\documentclass{eas}
\usepackage{graphicx}

\newcommand{\dvbump}{$\Delta{V}^{\mathrm{bump}}_{\mathrm{HB}}$}

\begin{document}
\title{From Theory to Observations and viceversa: theoretical uncertainties and observational constraints}   
\author{S. Cassisi}  
\address{INAF - Astronomical Observatory of Collurania, Via M. Maggini, 64100 Teramo, Italy, cassisi@oa-teramo.inaf.it} 

\runningtitle{S. Cassisi: Stellar model uncertainties and observational benchmarks}

\begin{abstract} 

During these last decades, our knowledge of evolutionary and structural properties of stars of different
mass and chemical composition is significantly improved. This result has been achieved as a consequence of our 
improved capability in understanding and describing the physical behavior of matter in the different
thermal regimes characteristic of the various stellar mass ranges and evolutionary stages.

This notwithstanding, current generation of stellar models is still affected by significant and, usually, not negligible
uncertainties. These uncertainties are related to our poor knowledge of some physical proceses occurring in the real stars
such as, for instance, some thermodynamical processes, nuclear reaction rates, as well as the efficiency of mixing processes.
These drawbacks of stellar models have to be properly taken into account when comparing theory with observations in order to derive
relevant information about the properties of both resolved and unresolved stellar populations. On the other hand, observations of both field
and cluster stars can provide fundamental benchmarks for constraining the reliability and accuracy of the theoretical framework.

In the following we review some important evolutionary and structural properties of very-low and low-mass stars, as well as the most important uncertainties affecting the stellar models for such stars. We show what are the main sources of uncertainty along the main evolutionary stages, and discuss the
present level of agreement between theory and observations.

\end{abstract}

\maketitle

\section{Introduction} 

During the second half of last century, stellar evolution theory has allowed us to 
understand the Color-Magnitude Diagram (CMD) of both Galactic globular clusters (GGCs) and open clusters, so that now 
we can explain the distribution of stars in the observed CMDs in terms of the nuclear evolution of
stellar structures and, thus, in terms of cluster age and chemical composition.
In recent years, however, the impressive improvements achieved for both photometric
and spectroscopic observations, has allowed us to collect data of an un-precedent accuracy, which provide at the same time a 
stringent test and a challenge for stellar models.

On the theoretical side, significant improvements have been achieved
in the determination of the Equation of State (EOS) of the stellar
matter, opacities, nuclear cross sections, neutrino emission rates,
that are the physical inputs needed in order to solve the stellar structure equations. At the same time, models computed with this updated physics have
been extensively tested against the latest observations. 

The capability of current stellar models to account for
all the evolutionary phases observed in stellar clusters is undoubtedly 
an exciting achievement  which crowns with success the development of 
stellar evolutionary theories. Following such a success, one is often tempted to use
evolutionary model predictions in an uncritical way, i.e., taking these results at
their face values without accounting for theoretical uncertainties. 
However, theoretical uncertainties do exist, as it is clearly shown by the
not negligible differences still existing among evolutionary results 
provided by the various research groups. 

The discussion of these uncertainties was early addressed by
Chaboyer (1995) in a pioneering paper investigating the reliability of  
theoretical predictions concerning H-burning structures presently evolving 
in GGCs, and in turn the accuracy of current predictions about GC ages. Such an investigation 
has been extended to later phases of stellar evolution by Cassisi et al.  (1998, 1999),  Castellani \& Degl'Innocenti (1999), Cassisi (2010, 2012),
and Vandenberg (2013); while Cassisi (2004)  and Cassisi \& Salaris (2011) have studied the issue of the main uncertainties affecting the evolutionary properties of 
intermediate-mass stars.

In the next sections, we will discuss  briefly the main \rq{ingredients}\rq\ necessary for
computing stellar models and show how the residual uncertainties on these inputs affect
theoretical predictions of the evolutionary properties of very-low and low-mass stars. Although not directly connected with 
the computation of stellar models, an other crucial ingredient in order to perform a suitable comparison between the theoretical
framework and observations is represented by the bolometric correction scale and a color-effective temperature relation for the various
photometric bands. We will show some shortcomings in present generation of color-${\rm T_{eff}}$ relations when discussing some comparisons 
between theory and observations. For a detailed discussion on the uncertainties affecting model atmospheres computations and, hence, the correlated 
color-${\rm T_{eff}}$ relations we refer the interested reader to Bessell~(2008) and Allard et al.~(2012) and references therein. For a detailed, recent analysis of the level of agreement between theoretical bolometric correction scales and empirical estimates we refer to Buzzoni et al.~(2010) and references therein.

\section{Stellar evolution: the ingredients} 
\label{ingredient}

The mathematical equations describing the physical behaviour of any stellar structure
are well known since long time, and a clear description of the physical meaning of each one 
of them can be found in several textbooks (as, for instance, Cox \& Giuli 1968, Kippenhan \& Weigart 1990, Salaris \& Cassisi 2005, and 
Cassisi \& Salaris 2013).

An accurate numerical solution of these differential equations is no longer a problem, and 
it can be easily and quickly achieved when using modern numerical solution schemes and current
generation of powerful computers. 
This notwithstanding, in order to solve these equations, boundary conditions have to be provided:
the boundary condition at the stellar centre are trivial (see Salaris \& Cassisi 2005); 
however the same does not apply for those at the bottom of the stellar atmosphere.
Let us briefly remember that \lq{to provide the outer boundary conditions}\rq\ means to provide the
values of temperature and pressure at the base of the stellar atmosphere: this requirement can be
accomplished either by adopting an empirical solar thermal stratification as provided by Krishna-Swamy (1966) 
 or a grey atmosphere when the so-called Eddington approximation is assumed (see section~\ref{rgb}).

A more rigorous procedure to obtain the outer boundary conditions is to use results from model atmosphere computations as early shown by Morel et al. (1994), and more recently by Salaris et al. (2002) and  Vandenberg et al. (2008). 
In general model atmospheres are computed considering a plane-parallel
geometry, and solving  the hydrostatic equilibrium equation together with the frequency dependent 
(no diffusion approximation is allowed in these low-density layers) equation of radiative
transport and convective transport when necessary, plus the appropriate equation of state.
In the following sections we will discuss the impact of different choices about the outer boundary conditions
on various theoretical predictions.

\subsection{The physical inputs} 

In order to compute a stellar model, it is fundamental to have an accurate description of the
physical behaviour of the matter in the thermal conditions characteristics of the stellar interiors
and atmospheres. This means that we need to know as much accurately as possible several physical
inputs as:

\begin{itemize}

\item{{\it the opacity}: the \lq{radiative}\rq\ opacity is related to the mean free path of photons
inside the stars and it plays a pivotal role in determining the efficiency of heat transfer via
radiative processes. When the stellar matter is under conditions of partial or full electron
degeneracy, electrons are able to transport energy with a large efficiency since they have a
longer mean free path than in case of non degenerate electrons. In this case, the energy transport
by conduction becomes quite important and the value of the conductive opacity has to be properly
evaluated.}

\item{{\it equation of state}: the EOS of the stellar matter is another key input for 
the model computations; it connects  pressure, density, temperature and chemical composition 
at each point within the star; determines the value of the adiabatic gradient 
(which is the temperature gradient in most of the
convective region), the value of the specific heat (which appears in the   
expression of the gravitational energy term), and plays a crucial role in the evaluation of    
the extension of the convective regions;}

\item{{\it nuclear cross-sections}: the  evaluation of the cross-sections for the various
nuclear burning processes occurring in the stellar interiors is mandatory in order to
properly establish the stellar energy budget. Even if thanks to laboratory experiments, many nuclear
cross-sections are nowadays known with a high accuracy, there are still some important
nuclear processes for both the H- and the He-burning, whose nuclear rate is poorly known;}

\item{{\it neutrino energy losses}: a precise determination of the energy losses due to 
neutrino emission is also important when the star is characterised by high density and low
temperature as it occurs in the interiors of Red Giant Branch (RGB) stars.}

\end{itemize}

It exists a quite rich literature describing the improvements which have been achieved 
concerning our knowledge of the physical inputs required for computing stellar models.
Therefore, in the following, unless quite relevant for our discussion, we will not discuss in detail this issue and
refer the interested reader to the exhaustive reference lists reported by Chaboyer (1995), Catelan et al. (1996), 
Cassisi et al. (1998, 1999, 2001), and Salaris et al. (2002).

\subsection{The microscopic mechanisms}

When computing a stellar model, some important assumptions have to be done concerning the
efficiency of some microscopic mechanisms. With the expression \lq{microscopic mechanisms}\rq\ 
we refer to all those mechanisms which, working selectively on the different chemical species, can
modify the chemical stratification in the stellar interiors and/or atmosphere. These mechanisms
are: atomic diffusion and radiative levitation.

\begin{itemize}

\item{{\it atomic diffusion}: it is a basic physical particle transport mechanism driven by collisions of
gas particles (Burgers~1969). Pressure, temperature and chemical abundance gradients are the driving
forces behind atomic diffusion. A pressure gradient and a temperature
gradient tend to push the heavier elements in the direction of
increasing pressure and increasing temperature, whereas the resulting
concentration gradient tends to oppose the above processes. The speed
of the diffusive flow depends on the collisions with the surrounding
particles. The efficiency of the different mechanisms involved in the atomic diffusion
process is given in terms of atomic diffusion coefficients which have to be estimated
on the basis of experimental measurements.}

\item{{\it radiative levitation}: it is an additional transport mechanism caused
by the interaction of photons with the gas particles, which acts
selectively on different atoms and ions (Michaud et al.~1976). Since
within the star a net energy flux is directed towards the
surface, photons provide an upward \lq{push}\rq\ to the gas
particle with which they interact, effectively reducing the
gravitational acceleration. Since, at the basis of this process there are the interactions
of photons with gas particles, it is clear that the efficiency of
radiative levitation is related to the opacity of the stellar
matter, in particular to the monochromatic opacity, and increases
for increasing temperature (radiation pressure $P_{rad}\propto{T^4}$). The
evaluation of the radiative accelerations is really a thorny problem due to the need of accounting
for various interaction processes between photons and chemical elements, and how the momentum of
photons is distributed among ions and free electrons.}

\end{itemize}

In the past, these non-canonical processes were usually ignored in stellar models
computations. However, helioseismology has clearly shown how important is to include atomic
diffusion in the computation of the so-called Standard Solar Model (SSM), in order to obtain a good
agreement between the observed and the predicted frequencies of the non-radial p-modes 
(Christensen-Dalsgaard, Proffitt \& Thompson 1993). 

However, although helioseismology strongly support SSMs accounting for atomic diffusion, 
recent spectroscopical measurements (Gratton et al. 2001, Ramirez et al. 2001, and references therein) of the
iron content in main sequence (MS) GGC stars, are in severe disagreement with the predictions provided by diffusive models: the measured
iron content does not appear to be significantly reduced with respect the abundance estimated for giant
stars in the same cluster as one has to expect as a consequence of diffusion being at work. In order to reconcile these two independent findings, one has to account for the fact that: i) radiative acceleration in the Sun can amount to about the 40\% of gravitational acceleration (Turcotte et al.~1998), and one can expect that its value is larger in more metal-poor, MS stars; ii) a slow mixing process below the solar
convective envelope could help in explaining better the observed Be and Li abundances (Richard et al.~1996)
and improve the agreement between the predicted and observed sound speed profile (Brun et al.~1999).
All these evidence seem to support the need to self-consistently account for both atomic diffusion, radiative levitation, and non canonical
mixing (at the bottom of the outer convection zone) in the computation of low-mass stellar models.

\subsection{The macroscopic mechanisms}

Being convection a non-linear, non-local, time dependent physical process, we still face
with the longstanding problem of how to manage the mixing processes in an
evolutionary code. To overcome this problem, the efficiency of convection is commonly treated by adopting some phenomenological
theory as discussed in the following discussion. 

The two issues that have to be solved when treating a convection zone in a stellar model are those related
to: {\it i)} the actual temperature gradient in such region, and {\it ii)} the real extension of the whole convective zone.
The first issue is really important only when considering the outer convective regions. This occurrence is due to the evidence that, in the stellar
interiors as a consequence of the high densities and, in turn, of the high capability of convective
bubbles to transport energy, the \lq{real}\rq\ temperature gradient is equal to the
adiabatic one. This consideration does not apply when considering the outer, low-density layers, where the correct 
temperature gradient has to be larger than the adiabatic one: the so-called {\sl superadiabatic gradient}. 

The adequate evaluation of this superadiabatic gradient is a thorny problem in low-mass stellar model computations. In this context, it
 is worth remembering that the radius and, in turn, the effective
temperature of cool stars (roughly speaking those stars with ${\rm T_{eff}<8000K}$) are drastically affected by the
choice of the superadiabatic gradient.

The equations still largely employed  to determine the value of the temperature gradient in 
the superadiabatic regions of stellar convective envelopes and atmospheres   are based on the so called ``mixing length theory'' (MLT).  The MLT in its original form is a simple, local, time independent model, firstly 
applied to stellar modelling by Biermann~(1932).  
Its description of convective transport considers the motion of 'average'
convective cells, all with the same physical properties at a given radial
location $r$ within the star. These convective cells have a mean size l 
(${\rm l=\alpha_{MLT} H_P}$ is the so-called 'mixing length', 
where ${\rm \alpha_{MLT}}$ is a free parameter and ${\rm H_P}$ is 
the local pressure scale height ${\rm H_P\equiv P/g \rho}$, the various symbols have their usual meaning) also equal to their
mean free path, and an average convective speed
${\rm v_c}$ at a given value of $r$.  There is in principle no reason why ${\rm \alpha_{MLT}}$ should be kept constant when considering  stars of different masses and/or chemical composition and/or at different evolutionary stages; even within
the same star ${\rm \alpha_{MLT}}$ might in principle vary from layer to layer, although
stellar models usually keep ${\rm \alpha_{MLT}}$ constant throughout the convective zone 
(but see Pedersen,  Vandenberg \& Irwin~1990, for some low-mass stellar 
models computed with varying ${\rm \alpha_{MLT}}$ within the convective envelope)§.
Keeping ${\rm \alpha_{MLT}}$ constant still means a change of the cells' mean size and free path due to
the variation of ${\rm H_P}$.

The value of ${\rm \alpha_{MLT}}$ is usually calibrated by reproducing some empirical
constraint. In case of stellar models, it is typically the calibration\footnote{A SSM is a stellar evolution model that best reproduces the observed properties of the Sun. The fundamental observational properties that any SSM must satisfy are the solar mass value ${\rm M_\odot = 1.989\times10^{33}}$~g, the surface luminosity ${\rm L_\odot = 3.486\times10^{33}}$~erg/s, and radius ${\rm R_\odot = 6.9599\times10^{10}}$~cm, at the solar age ${\rm t_\odot = 4.57}$~Gyr. In addition, the predicted surface chemical composition should be consistent with the observed photospheric composition.} of a SSM
that determines the value of ${\rm \alpha_{MLT}}$. In all stellar model databases  (Girardi et al.~2000, Vandenberg et al.~2000, Yi et al.~2001, Pietrinferni et al.~2004, Dotter et al.2007) this solar calibrated $\alpha_{MLT}$ is kept fixed in the evolutionary 
calculations of stars of different masses and chemical compositions.

There are 3 additional free parameters entering the MLT  
that are generally fixed a priori, before the calibration of ${\rm \alpha_{MLT}}$. 
They will be denoted here as $a, b, c$ following the formalism 
of Tassoul et al.~(1990). These three parameters plus ${\rm \alpha_{MLT}}$ enter the equations that determine 
${\rm v_c}$, the average convective flux 
$F_c$ and the convective efficiency $\Gamma$ (defined as 
the ratio between the excess heat content of a raising convective bubble just before its 
dissolution, and the energy radiated during its lifetime)
at each point within the convective envelope. Once $a, b, c$ are fixed, the temperature gradient at a 
given location in the convective envelope of 
a stellar model and the resulting ${\rm T_{eff}}$, depend only on the value of ${\rm \alpha_{MLT}}$.
On the other hand, when ${\rm \alpha_{MLT}}$ is kept fixed and one or more of the other three parameters 
are varied, the temperature gradients and ${\rm T_{eff}}$ are also affected (see Henyey et al.~1965).

The \lq{classical}\rq] formulation of the MLT by B{\"o}hm-Vitense~(1958) that is nowadays almost 
universally used in computations of  stellar evolution models (hereinafter the BV58 'flavour' of the MLT) 
employs a specific set of choices for the values for $a, b$ and $c$. 
Solar calibrations with the present generation of stellar input physics  provide a value of ${\rm \alpha_{MLT}}$ typically around $\sim$2.0,  the precise value (for a given set of input physics) being dependent on the choice for the outer boundary conditions. There is another flavour of the MLT, commonly employed in calculations of model atmospheres for white dwarf stars (Rohrmann~200). 
It is the so-called ML2 (Tassoul et al.~1990), with its own specific choice of $a, b$ and $c$, 
different from the case of BV58. However, Salaris \& Cassisi (2008) have shown that the two flavours for the MLT
provide quite similar results concerning the ${\rm T_{eff}}$ scale of stellar models, once the values of 
${\rm \alpha_{MLT}}$ have been properly calibrated by computing a SSM.

It is worth recalling that there exists also an alternative formalism for the computation 
of the superadiabatic   
gradient, which in principle does not require the calibration of any free   
parameter. It is the so-called Full-Spectrum-Turbulence theory (FST,   
Canuto \& Mazzitelli~1991, Canuto, Goldman \&   
Mazzitelli~1996), a MLT-like formalism with a more sophisticated   
expression for the convective flux, and the scale-length   
of the convective motion fixed a priori (at each point in a convective region,   
it is equal to the harmonic average between the distances from the top and the bottom convective   
boundaries). From a practical point of view, the FST theory contains also a free parameter
which has to be fixed, even if it seems to have a physical meaning larger than that
of ${\rm \alpha_{\rm MLT}}$.

The problem of the real extension of a convective region really affects both convective core and
envelope. In the canonical framework it is assumed that the border of a convective region is fixed
by the condition - according to the classical Schwarzschild criterion - that the radiative gradient
is equal to the adiabatic one. However, it is clear that this condition marks the point where the
acceleration of the convective cells is equal to zero, so it is realistic to predict that the
convective elements can move beyond, entering and, in turn, mixing the region surrounding the
classical convective boundary. This process is commonly referred to as convective overshoot.

Convective core overshoot is not at all a problem for low-mass stars during the H-burning stage
since the burning  process occurs in a radiative region. However, the approach used for treating
convection at the border of the classical convective core is important during the following
core He-burning phase as discussed in the section~\ref{hestage}. Convective envelope overshoot could be
important for low-mass stars, since these structures have large convective envelope during the
shell H-burning phase and the brightness of the bump along the RGB could be significantly affected
by envelope overshoot (see section~\ref{bump}).

\section{Very low-mass stars}
\label{vlm}

The interest in the study of physics at work in objects at the bottom of and below the MS dates back to the early demonstration made by Kumar~(1963) that it has to exist a minimum total mass allowing a star to ignite hydrogen, the so called \lq{H-burning minimum mass}\rq\ (HBMM), and that below this critical mass, hydrostatic equilibrium is guaranteed by pressure provided by degenerate electrons. An additional historical reason for which low-mass and very low-mass stars (VLM) have been very important objects in the stellar astrophysics field is related to the fact that, before the results from microlensing surveys such as MACHO and OGLE,  were published, they were considered together with white dwarfs and sub-stellar objects, the main contributors to the dark matter budget in the Galaxy. Nowadays, there is a renewed interest in these stellar structures due to the fact that they are suitable candidates for the search of exoplanets. 

\subsection{The physical properties}

The thermodynamical conditions experienced by stellar structures in the VLM regime are very extreme. Just in order to have an idea, the central temperature that in the Sun is  $\sim10^7$~K is of the order of $4\times10^6$~K for a VLM star with mass equal to about the HBMM value - i.e. $\sim0.1M_\odot$ at solar chemical composition -, while the central density does increase from ${\rm \sim100g cm^{-3}}$ for the Sun to ${\rm \sim500g cm^{-3}}$ in a ${\rm 0.1M_\odot}$ VLM star. In the same mass range, the temperature at the basis of the photosphere ranges from $\sim6000$~K to $\sim2500$~K when decreasing the stellar mass, while the density at the same location spans a range from ${\rm \sim10^{-7}g cm^{-3}}$ to  ${\rm \sim10^{-5}g cm^{-3}}$.

Under these thermal conditions, the coupling parameter $\Gamma$ used for indicating how much the Coulomb interactions among ions are important is in the range between 0.1 - 30, which means that non ideal, Coulomb interactions among ions are quite important in the evaluation of the EOS. When the density becomes quite large one has also to consider the occurrence of pressure dissociation and ionization. This condition is well achieved in the VLM regime and so these processes have to be also accounted for, as well as in the EOS is important to consider the process of formation/dissociation of molecules - mostly the ${\rm H_2}$ one.

The electron degeneracy parameter $\eta$ in these peculiar objects is of the order of unity and it increases significantly when decreasing the stellar mass, so this implies that in the lowest mass tail of the VLM range there are conditions of partial electron degeneracy.

On the basis of these evidence, we can describe a VLM structure as an object where molecular H, atomic helium, and many other molecules (see below) are stable in the atmosphere and in the outer envelope layers, while the stellar interiors are mainly formed by a fully ionized H/He plasma; Coulomb interactions are important in the whole mass regime as well as the pressure ionization/dissociation process, while electron degeneracy increases with decreasing mass and becomes quite important for a mass of about ${\rm 0.12M_\odot}$. It is evident that, due to the complex thermodynamical conditions present in these objects, the evaluation of a reliable EOS has been for long time a thorny problem, until the release of an accurate EOS for cool and dense stars by Saumon et al.~(1995).

An important issue for VLM stars is related to the complicated task of computing accurate model atmospheres which are indeed important, not only in order to allow a description of the emergent radiative flux - so allowing to provide model predictions about  colors and magnitudes in several photometric planes -, but also for fixing the outer boundary conditions needed for solving the stellar structure equations (see below).
The huge difficulties in the VLM model atmosphere computations arise, once again from the peculiar physical properties which characterize these objects. In fact, due to the cool temperatures and high pressures of the outer stellar layers, the opacity evaluations is made extremely complicated by the presence of a huge numbers of molecules whose individual contributions to the Rosseland mean opacity has to be properly evaluated.

By simply looking to the spectrum of a VLM star (see Fig.~2 in Allard \& Hauschildt~1995) one easily realize that, around and below ${\rm T_{eff}\approx4000}$K, the presence of molecules such as ${\rm H_2O}$, CO, VO and TiO is very important. More in detail, TiO and VO governs the energy flux in the optical wavelength range, while ${\rm H_2O}$ and CO are the dominant opacity source in the infrared spectral window. Then, when the effective temperatures attain values lower than ${\rm\sim2500-2800}$\,K, the process of grains condensation becomes extremely important, so increasing of orders of magnitude the difficulty of a reliable opacity and, hence, model atmosphere computations.

In addition, VLM stars share with white dwarfs an important opacitive properties: due to the large density and pressure in the outer layers the Collisional Induced Absorption (CIA) on ${\rm H_2}$ molecules becomes a quite important contributor to the radiative flux absorption. Since the ${\rm H_2}$  molecule in its ground electronic state has no electric dipole, absorption of photons can take place only via electric quadrupole transition. This is the reason for which low-density molecular hydrogen gas is essentially transparent throughout the visible and infrared portion of the spectrum. However, when the density is large enough, each time a collision between two particles occurs, the interacting pair such as ${\rm H_2-H_2}$, ${\rm H_2-He}$ or ${\rm H_2-H}$, forms a sort of \lq{virtual molecule}\rq\ which, because of its nonzero electric dipole, can absorb photons with a probability which is much higher than that of an isolated ${\rm H_2}$ molecule.

When ${\rm H_2}$ CIA is efficient - as it occurs for stars with mass lower than about ${\rm \sim0.2M_\odot}$ - it suppresses the flux longward of ${\rm 2\mu{m}}$ and causes the redistribution of the emergent radiative flux towards shorter wavelengths, i.e. for suitable choices of the photometric bands VLM stars tend to appear \lq{bluer}\rq\ with decreasing mass (Allard et al.~1997).

As already mentioned, an additional problem related to the computation of VLM stellar models is the determination of accurate outer boundary conditions. Although for low-mass stars it is a common procedure to adopt the outer boundary
conditions provided  by a grey model atmosphere - and usually this assumption provide results in quite good agreement with those based on more accurate and sophisticated model atmospheres (see section~\ref{rgb}) -, {\sl in principle} for VLM stars this approach can not be adopted. In fact, the \lq{grey model atmosphere}\rq\ approximation is valid only when all the following conditions are fulfilled: i) presence of an isotropic radiation field, ii) radiative equilibrium, iii) the radiative absorption is independent on the photon frequency. However, in VLM stars the strong frequency-dependence of the molecular absorption coefficients yields synthetic spectra which depart severely from a frequency-averaged energy distribution, and so the last condition is not clearly satisfied in these structure. 

More importantly, below $\sim 5000$ K, molecular hydrogen recombination in the envelope (H+H $\rightarrow$ H$_2$) reduces the entropy and thus the adiabatic gradient. This occurrence, coupled with the large radiative opacity and, hence, large value of the radiative gradient (${\rm \nabla_{rad}\propto\kappa}$), favors - according to the Schwarzschild criterium - the presence of a convective instability in the atmosphere, so that convection penetrates deeply into the optically thin layers (Baraffe et al.~1995), so radiative equilibrium is no longer satisfied.
All these evidence provide compelling arguments supporting the idea that, in order to provide a physically grounded description of the sub-atmospheric and envelope layers of VLM structures, boundary conditions provided by accurate, non-grey model atmospheres have to be adopted (Chabrier \& Baraffe~2000). Usually these model atmosphere predictions are \lq{attached}\rq\ to the interior solution at a value of the optical depth $\tau\approx100$, because at this depth the diffusive process approximation is fully fulfilled. 

\begin{figure}
\begin{centering}
\vskip -0.5cm
\includegraphics[width=9.5cm]{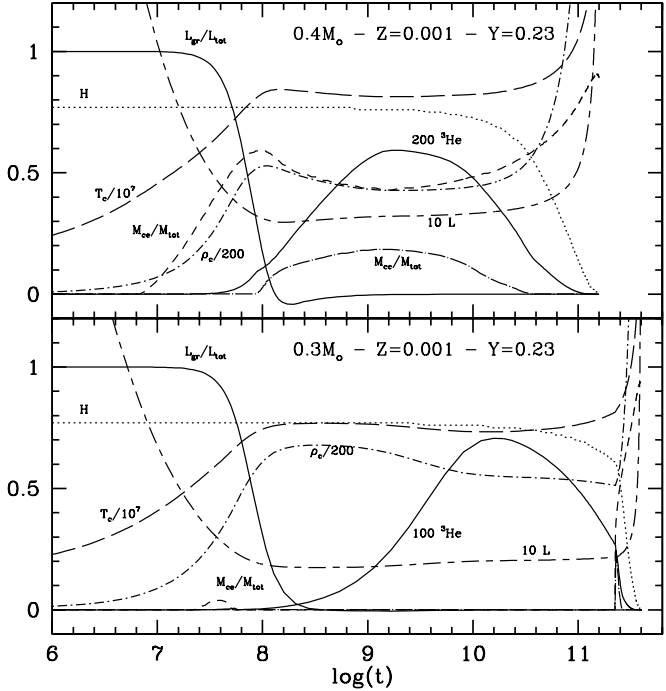}
\vskip -0.3cm
\caption{The temporal evolution of the surface luminosity (L) in ${\rm L_\odot}$, the ratio between the gravitation energy and the total luminosity (${\rm L_{gr}/L_{tot}}$), central density (${\rm \rho_c}$), central temperature (${\rm T_c}$), size of the convective core (${\rm M_{cc}}$) as fraction of the total mass, mass location of the bottom of the convective envelope (${\rm M_{ce}}$), and central abundances by mass of H and ${\rm ^3He}$ for two models at the transition between fully convective stars and those with a radiative core, i.e. a ${\rm 0.4M_\odot}$ and ${\rm 0.3M_\odot}$ models, namely.  }
\end{centering}
\label{strutvlm}
\end{figure}

A quite important, and exclusive (when not considering the  pre-MS stage), characteristic of VLM structures is that below a certain mass limit, they are fully convective objects. This is due to two concomitant processes: i) the large radiative opacity with characterizes both the interiors and the outer layers, which
causes a huge increase of the radiative gradient, ii) the decrease in the envelope layers of the adiabatic gradient due to the formation of molecules. The transition mass between fully convective stars and structure with a radiative core is around ${\rm \sim0.35M_\odot}$, the exact value depending on the metallicity. 

It is also important to note that convection in VLM stars is largely adiabatic since the large densities make the heat convective transport quite efficient. This represents a significant advantage of VLM stars with respect more massive objects, because model predictions do not depend at all on the value adopted for the free parameter - the mixing length - entering in the mixing length theory. On the other hand, the fact that VLM have extended convective envelope or are fully convective has an important implication: the depth of the convective envelope or - for fully convective stars - the global thermal stratification of the structure is affected by the choice of the outer boundary conditions (see the discussion in Salaris \& Cassisi~2005). So, the choice of how to fix the boundary conditions at the bottom of the atmospheric layers does not affect only the ${\rm T_{eff}}$ predictions but also the global structural properties.

As far as it concerns the thermonuclear H-burning processes important for the energy budget in VLM stars, due to the low central temperatures which characterize their interiors the nuclear reactions that are really important are:
${\rm p + p \rightarrow D + e^+ + \nu_e}$ and ${\rm p + D \rightarrow {^3He} + \gamma}$.

The process of destruction of ${\rm ^3He}$, ${\rm ^3He(^3He,^4He)2p}$ is really effective only for ${\rm T > 6\times10^6}$~K, this means that
${\rm ^3He}$ behaves as a pseudo-primary element for a large fraction of the whole core H-burning stage. This occurrence has the important consequence that the definition of the \lq{Zero Age Main Sequence}\rq\ for VLM stars is largely meaningless, because they live a huge fraction of their H-burning lifetime without attaining the equilibrium configuration for ${\rm ^3He}$. The characteristic timescale for this process for stellar structures with a metallicity ${\rm Z=10^{-3}}$
is of the order of $\sim15.8$~Gyr for a ${\rm 0.4M_\odot}$, and $\sim126$~Gyr for a ${\rm 0.15M_\odot}$, while for the same structures the core H-burning lifetime is $\sim158$~Gyr and $\sim1260$~Gyr respectively, i.e. larger than the Hubble time.
\begin{figure}
\begin{centering}
\vskip -2.5cm
\includegraphics[width=11cm]{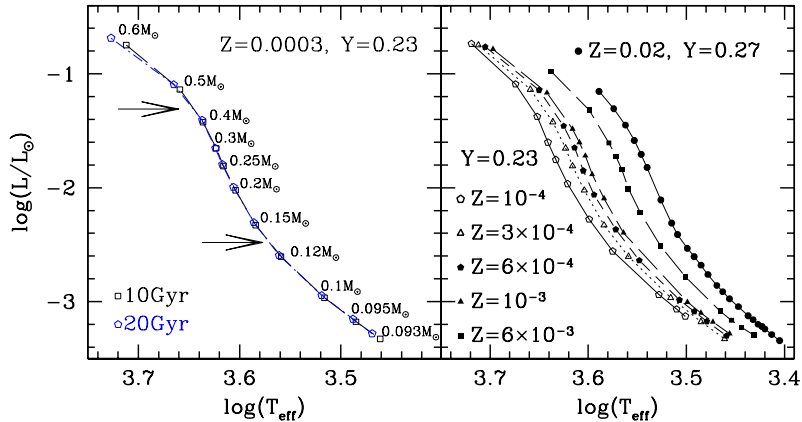}
\vskip -3.5cm
\caption{{\sl Left panel}: The HR diagram of low-mass and VLM stars for two different age assumptions, and for Z=0.0003, Y=0.23. The arrows mark the location of the two bending points discussed in the text. {\sl Right panel}: as left panel but for various metallicities.
}
\end{centering}
\label{hrvlm}
\end{figure}

\subsection{The structural and evolutionary properties}

Figure~1 shows the time behavior of selected structural parameters for two VLM models with mass equal to 0.4 and ${\rm 0.3M_\odot}$, respectively. The ${\rm 0.4M_\odot}$ model behaves like models with moderately larger masses populating the upper portion of the MS. The increase of the ${\rm ^3He}$ abundance toward its equilibrium value increases the efficiency of the H-burning\footnote{The {\sl p-p} chain is more efficient when it achieves the equilibrium of the secondary elements.} and the structure reacts decreasing both central temperature and density, which start increasing again only	when the equilibrium value for	${\rm ^3He}$ has been attained. However, fully convective structures as the ${\rm 0.3M_\odot}$ model behave quite differently, and central density keeps decreasing all along the major phase of H burning. One finds that the central density of such a model starts suddenly increasing again in the very last phases of central H burning, when the increased abundance of He and the corresponding decrease of radiative opacities induces a radiative shell which rapidly grows to eventually form a radiative core.

Left panel of fig.~2 shows the effect of age on the HR diagram of a set of models. One can easily note that the age plays a negligible role on the H-R diagram location of stars below about ${\rm 0.5M_\odot}$. It is curios that the ${\rm 0.3M_\odot}$ model appears the less affected by age, less massive models shoving a progressively increasing sensitivity to the adopted ages. One can understand the reason for such a behaviour by considering - as shown in fig.~1 - that the ${\rm 0.3M_\odot}$ model, at t=10Gyr, has already achieved at the centre the ${\rm ^3He}$ equilibrium configuration,  so that the following evolution is governed by the depletion of central H only (which has a very long timescale). On the contrary, less massive models keep increasing the central abundance of ${\rm ^3He}$ - an occurrence that affects the {\sl p-p} chain efficiency - and readjusting the structure according to such an occurrence. A readjustment that in the less massive models is also amplified by the decreasing electronic degeneracy induced by the decrease of central density.
\begin{figure}
\begin{centering}
\vskip -2.8cm
\includegraphics[width=12cm]{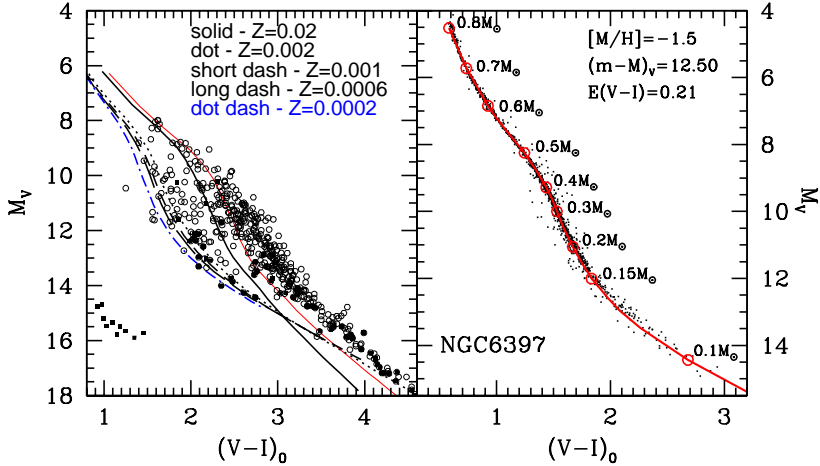}
\vskip -3.7cm
\caption{
{\sl Left panel}: The optical CMD for field stars with known parallaxes (see Cassisi et al.~2000, for more details), with superimposed theoretical models for the labeled values of the metallicity. The two solid lines refer to the same solar metallicity models transferred in the observational plane by using two different colour - ${\rm T_{eff}}$ relations (see text for more details). {\sl Right panel}: The HST CMD of the MS locus of the GGC NGC~6397 compared with VLM models for the appropriate metallicity.
}
\end{centering}
\label{fig:vvi_vlm}
\end{figure}
The right panel of fig.~2 shows the H-R diagram of VLM structures with an age of 10~Gyr and for some selected metallicities: as a consequence of increased opacity when the metallicity increases, the H-R locus becomes cooler and fainter with increasing heavy elements abundance. However, we wish to note that the MS locus of VLM stars shows two well-defined bending points as indicated by the arrows in the left panel of the same figure:

\begin{itemize}

\item{the brighter point is located at ${\rm T_{eff}\approx4500}$~K, and corresponds to an evolutionary mass of ${\rm \sim0.5M_\odot}$: the physical process at the origin of this change in the MS slope is the molecular hydrogen recombination. With decreasing total mass, at a certain point the outer stellar layers achieve the thermal conditions which make the ${\rm H_2}$ formation process quite efficient. This occurrence causes a decrease of the number of free particles in the plasma, and hence a decrease of the adiabatic gradient to
${\rm \nabla_{ad}\sim0.1}$ (for the sake of comparison, a perfect, mono-atomic gas has a value ${\rm \nabla_{ad}\sim0.4}$). The decrease of the adiabatic gradient induced by ${\rm H_2}$ recombination produces a flatter temperature gradient and so the effective temperature of the model is larger with respect a model where this process is not accounted for (Copeland et al.~1970). When decreasing the metallicity, this bending point shifts at larger ${\rm T_{eff}}$ values, i.e. larger mass, because the outer layers of metal-poor stars are denser\footnote{Let us remember that, the lower the metallicity, the lower the opacity, and since ${\rm dP/d\tau=g/\kappa}$, for a fixed gravity, at the same optical depth the pressure is larger.}, an occurrence that favors the formation of molecular hydrogen. Since the location of this point in the H-R diagram strongly depends on the thermal properties of the stellar envelope, its comparison with suitable observational constraints represents a formidable benchmark of the reliability of the adopted EOS;}

\item{the faintest bending point is located at ${\rm T_{eff}\approx2800}$~K and corresponds to an evolutionary mass of ${\rm \sim0.15M_\odot}$. Its presence is due to the fact that decreasing the total mass below this critical value the level of electron degeneracy does increase. As a consequence, the contribution of degenerate electron pressure to the total pressure increases, and eventually this will produce the well-known mass-radius relation for fully degenerate objects, i.e. ${\rm R \propto M^{-1/3}}$. On the basis of previous discussion, one can now predict that, when decreasing the metallicity, this point \lq{shifts}\rq\ toward larger effective temperature, i.e. larger stellar mass. Clearly, the location of this point in the HR diagram is more affected by the opacity evaluations which control the outer layers pressure stratification than by the EOS.}

\end{itemize}
\begin{figure}
\vskip -3.5cm
\begin{centering}
\includegraphics[width=12.5cm]{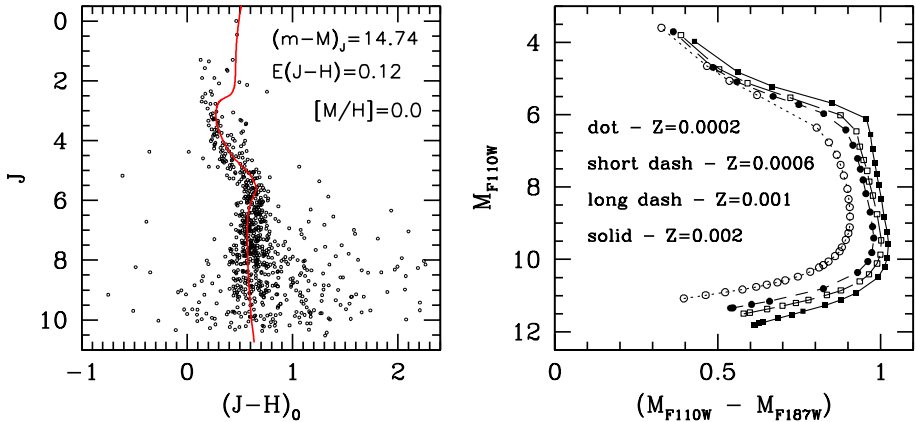}
\vskip -4.0cm
\caption{{\sl Left panel}: The near-infrared CMD of the stellar population in a window of the Galactic bulge (Zoccali et al.~2000) compared with models for solar metallicity. {\sl Right panel}: The CMD in the F110W and F187W NICMOS HST filters  for stellar models with mass ${\rm \le0.8M_\odot}$, and for selected metallicities.
}
\end{centering}
\label{fig:infrared_vlm}
\end{figure}

We have already defined the HBMM quantity, i.e. the minimum stellar mass that can attain thermal equilibrium supported by H-burning. Detailed numerical computations (Chabrier \& Baraffe~1997, 2000) based on the most accurate physics for VLM stars have shown that for solar metallicity the HBMM is equal to ${\rm 0.075M_\odot}$. The value of HBMM does increase with decreasing metallicity, being equal to ${\rm 0.083M_\odot}$ for Z=0.0002. This is due - as already discussed - to the fact that, at fixed total mass,  decreasing the metallicity, the opacity decreases and the stellar structures become more and more dense so increasing the level of electron degeneracy, a process that opposes to the achievement of the thermal conditions required for an efficient ignition of H-burning.

\subsection{The observational properties}

\subsubsection{The Color-Magnitude diagrams}

The location of the MS loci for VLM structures in an optical CMD is shown in fig.~4, and compared with some empirical measurements for both field dwarf stars with well known parallaxes, and a GGC. It is worth noting that, while the theoretical sequences reproduce finely both the location and shape of the observed MS in the metal-poor regime, this is not true when considering solar metallicity, field dwarfs. 
The solar metallicity models have been transferred from the H-R diagram to the observational plane by using two different color-${\rm T_{eff}}$ relations:
the calibration provided by  Allard \& Hauschildt~(1995, thin solid line) and the more updated one by Allard et al.~(1997, thick solid line). This experiment has been done in order to show how much crucial (and critical) the accuracy of the color-${\rm T_{eff}}$ relation is in comparing theory with observations in the VLM star regime. The existence of a significant disagreement between solar metallicity stellar models, based on more accurate physics presently available,
and empirical data in optical bands, represents a plain evidence for the presence of some shortcoming in the available color-${\rm T_{eff}}$ relation for the optical photometric bands in the metal-rich regime\footnote{The problem in the optical bolometric corrections is probably associated to specific radiative opacity contributions at wavelengths shorter than ${\rm \sim1\mu{m}} $, that are not properly accounted for in model atmosphere computations. There are some indications that the \lq{missing opacity contribution}\rq\ could be associated to the TiO line list with some contribution coming also from other molecules such as CaOH.}. In fact, the same stellar models which are not able to match the data shown in the right panel of fig.~4, match pretty well observations in the near-infrared bands as shown in the right panel of fig.~5.

\begin{figure}
\begin{centering}
\vskip -3.5cm
\includegraphics[width=10cm]{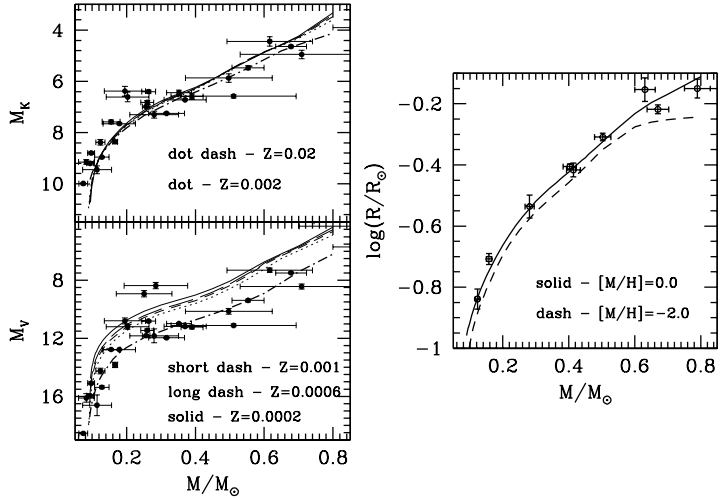}
\vskip -0.8cm
\caption{
{\sl Right panel}: theoretical mass - luminosity relations in different photometric bands and for various metallicities. The empirical data correspond to measurements for field stars (Henry \& McCarthy~1993). {\sl Left panel}: the mass - radius relation for low-mass and VLM structures. Empirical data are from 
S{\'e}gransan et al.~(2003).}
\end{centering}
\label{fig:maslum_vlm}
\end{figure}

It is important to note that the morphology of the MS locus for VLM stars changes significantly when moving from the optical photometric planes to the near- and far-infrared ones, as shown in fig.~5. The most interesting evidence is that, from a few magnitudes below the MS Turn-off, the MS runs almost vertical in the NIR CMDs before suddenly shifting toward bluer colors with decreasing stellar mass. This occurrence is related to the ongoing  competition - in these infrared colors - between the tendency of the stellar color to become redder, due to the decreasing effective temperature and increasing radiative opacity in the optical, and the increasing CIA of ${\rm H_2}$ in the infrared which shifts back the flux to shorter wavelengths. This process leads to quasi-constant color sequences from a mass of ${\rm \sim0.5M_\odot}$, corresponding to ${\rm T_{eff}\approx 4500}$~K, for which the molecular hydrogen recombination starts being efficient to ${\rm \sim 0.1M_\odot}$. For stellar masses below this limit, the second effect becomes the dominant one as a consequence of the larger densities (${\rm H_2}$ CIA is proportional to the square power of the density), and this produces the blue loop at the very bottom of the MS. The blue loop becomes still more evident with decreasing metallicities, due to the larger density which characterizes the atmosphere of metal-poor stellar structures.

\subsubsection{The mass-luminosity and the mass - radius diagrams}

Figure~6 shows the mass-luminosity relations for selected photometric bands ranging from the visual V band to the near-infrared K band. One can easily note that the mass-luminosity relation becomes largely insensitive to the stellar metallicity when going from the visual to the near-IR photometric bands, with the ${\rm mass - M_K}$ relation almost unaffected by the metallicity for mass below ${\rm \sim0.4M_\odot}$. This occurrence is due to the fact that below ${\rm T_{eff}\sim 4000-4500}$~K, the opacity in the optical bands (dominated by TiO and VO) increases with metallicity so that the peak of the energy distribution is shifted toward larger wavelengths, in particular to the K band. This process causes a decreases in the flux emitted in the V band and conversely an increase in the K band with increasing metallicity. 

On the other hand, for a given mass, ${\rm T_{eff}}$ decreases with increasing metallicity, so that the total flux (${\rm F \propto T_{eff}^4}$) decreases. In the K band the two effects compensate, and hence the mass - luminosity relation in the K band is almost independent on the heavy elements abundance. However, this trends holds until the ${\rm H_2}$ CIA does not suppress largely the flux emitted in the K band, as it occurs in the more metal-poor VLM stars with mass just above the HBMM.

In passing, we note that the mass - magnitude relations show also some inflection points, clearly associated to the same physical processes producing the bending points in the H-R diagram. The presence of these points has to be properly taken into account when these mass - magnitude relations are used for retrieving the Initial Mass Function (IMF). In fact, in deriving the IMF the really crucial ingredient is the prime derivative of these relations with respect the stellar mass: since in correspondence of these inflection points the derivative shows a maximum, an accurate evaluation of this function is mandatory in order to avoid shortcoming in the IMF evaluation.

Thanks to the improvements in the observational facilities, the radii of many VLM stars have been determined accurately. The study of eclipsing binaries, interferometric measurements with the Very Large Telescope Interferometer, and transit observations from microlensing surveys have provided a large sample of reliable radii and mass measurements for VLM and low-mass stars. Figure~6 also shows a comparison between some theoretical predictions and observed radii for stars in the range of interest: a quite good agreement seems to exist and this evidence provides a sound support to the reliability and accuracy of last generation of stellar models for VLM structures. This notwithstanding, there are some indications (see Torres et al.~2010, and references therein) that stellar models could really underestimate the stellar radius and effective temperature for masses below the solar one with respect observational data for detached eclipsing binaries: the difference in the measured radius and ${\rm T_{eff}}$ with respect model predictions being $\sim5-10$\% and $\sim5$\%, respectively.
The radius inflation is usually explained as the consequence of an enhanced level of magnetic activity, induced by the high rotational rate of the stars in the binary systems (due to a spin-orbit synchronisation produced by tidal interactions). Actually the presence of a strong magnetic field can significantly affect the structural properties of VLM stars as suggested by Mullan \& MacDonald~(2001), and this occurrence can modify the predicted radii and effective temperature scale. For a recent reanalysis of this issue we refer to Spada et al.~(2013), although it is clear that additional theoretical and observational investigations are mandatory.

\section{Core H-burning stars}
\label{coreh}

\subsection{The location of the MS locus and the MS fitting technique}
\label{mslocation}

The comparison of theoretical isochrones at a given metallicity (and He content) shows that 
the brightness of the lower MS is typically unaffected by age. If one considers an upper limit to 
stellar ages equal to 14~Gyr, magnitudes and colours of the MS at ${\rm M_V}$ larger than $\sim$5.0 -- 5.5 
are age independent, and are affected only by the initial chemical composition (see left panel of fig.~7). 
These properties are exploited by the MS-fitting method (Salaris~2012, and references therein), widely employed to determine star cluster distances.
In case of clusters whose CMD morphology clearly identify as 'young', 
brighter sections of the MS can be employed, and this allows to apply the MS-fitting technique to reach distances 
outside the Galaxy for instance, clusters in the Large Magellanic Cloud.

A representation of the MS fitting technique is shown in fig.~6. One needs to determine 
observationally a deep CMD of the target population, and consider a template MS of known 
absolute magnitudes and dereddened colours, with the same chemical composition  -- that we label here in terms of 
[Fe/H], assuming a universal metal distribution and helium-enrichment ($\Delta$Y/$\Delta$[Fe/H]) law --  of the target. 
The difference between the absolute magnitudes of the template MS and the apparent magnitudes 
of the observed one (after correcting for the effect of reddening) provides the population distance modulus. 

\begin{figure}
\begin{centering}
\vskip -0.5cm
\includegraphics[width=7cm]{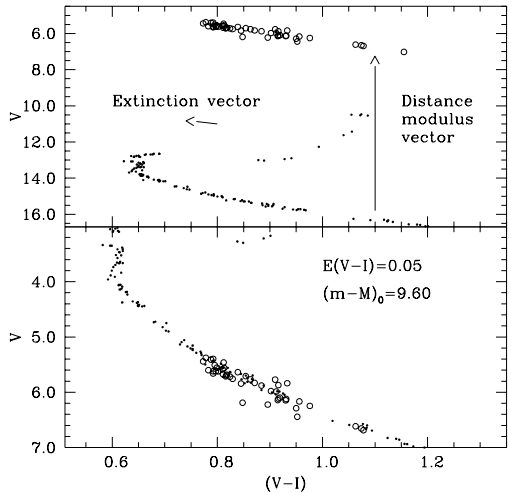}
\vskip -0.5cm
\caption{A representation of the MS-Fitting method applied to 
the Galactic open cluster M67. The template MS made of local stars with accurate parallaxes is displayed as 
large open circles. M67 CMD is denoted by small filled circles.} 
\label{MSfitt1}
\end{centering}
\end{figure}

To determine accurate distances, anything that may introduce systematic differences 
between intrinsic and observed colours 
and magnitudes of the target and the template MS must be accounted for.
For a detailed discussion about the uncertainties affecting the MS fitting-based distances to GGCs associated with parallax errors and associated biases (Lutz-Kelker and Malmquist bias) in the template MS stars, field binary 
and cluster binary contamination, reddening scale and photometric calibration we refer to the analysis performed by Gratton et al.~(2003)
The following discussion will focus on the accuracy of the template MS, and all additional source of potential 
uncertainty related with the theoretical evolutionary framework. 
 
In fact, in principle one could use as template the MS of theoretical isochrones calculated 
for the specific [Fe/H] of the target population. This choice is however hampered by 
the existing uncertainties in  the ${\rm T_{eff}}$ scale of stellar models with convective envelopes, and colour transformations (see right panel of fig.~7). It is
worth remembering that an uncertainty of only 0.02~mag in colours typically translates into a $\sim$0.10~mag error in the derived distance modulus,  
because in the widely employed ${\rm V-(B-V)}$ and ${\rm V-(V-I)}$ CMDs the MS slope is equal to 
$\sim$5.0--5.5 at ${\rm M_V}$ between $\sim$5.0 and $\sim$8.0, the typical magnitude range employed to determine 
distances to old populations. 

\begin{figure}
\begin{centering}
\vskip -2.5cm
\includegraphics[width=10cm]{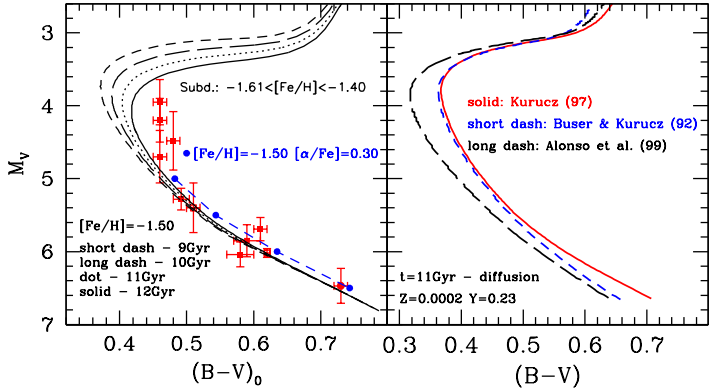}
\vskip -2.5cm
\caption{{\sl Left panel}: comparison between data for subwarfs with well-know parallax and metallicity in a specific range and scaled solar, theoretical
isochrones for various assumptions about the age and a metallicity appropriate for the selected subdwarf sample. For the sake of comparison, the fainter portion of the MS locus for an $\alpha-$enhanced, isochrone for the same iron content, is also shown. {\sl Right panel}: the CMD location of a theoretical isochrone transferred from the theoretical plane to the adopted observational one by using independent colour - ${\rm T_{eff}}$ relations.} 
\label{MSfitt2}
\end{centering}
\end{figure}

Right panel of fig.~7 shows a comparison between suitable theoretical isochrones and empirical data for local subdwarfs with known parallax and metallicity: it appears a fine agreement between the location of the theoretical MS loci
and the observational data.

This notwithstanding, the standard method is to build an empirical MS by considering samples of 
field dwarfs (or subdwarfs) of 
known [Fe/H] with accurate parallax distances and negligible (or well known) reddening. The main difficulty with this 
approach is that -- given the current sample of stars with accurate parallaxes and spectroscopically determined 
[Fe/H] -- only few objects will generally have the exact [Fe/H] of the target population, making difficult to 
determine an appropriate template MS. To overcome this problem one needs to shift the position of many  
field stars of varying [Fe/H] -- in reasonably narrow 
ranges around the target metallicity -- to the location they would have at the [Fe/H] of the target population. 
The effect of varying [Fe/H] is to 
change the colour and magnitude of stars 
of fixed mass along the unevolved MS; i.e. magnitude and colours at fixed mass increase when the metal content increases.
Adjustments to both magnitude and colour 
would therefore be necessary to match each field object to a star of equivalent mass at the metallicity of the cluster.
However, the slope of the MS -- at least in the 
${\rm M_V}$ range between $\sim$5.0 and $\sim$8.0 -- appears to be nearly independent of [Fe/H] in the 
metallicity regime of GGCs, and in the typical metallicity range 
of open clusters, so that templates constructed using colour shifts only are a reliable choice.

Once again, in principle one could estimate the required metallicity dependent colour shifts by relying on the theoretical framework, but
one has always to keep in mind the uncertainty in the adopted colour - ${\rm T_{eff}}$ relation, and the fact that the ${\rm T_{eff}}$ scale of MS stellar models is also affected by the physical inputs used in evolutionary computations. Left panel of fig.~\ref{MSfitt2} shows how huge can be the impact of using different prescriptions about the colour - ${\rm T_{eft}}$ scale on the location of the predicted MS locus in the observational plane. In this context, however one has to keep in mind that in evaluating the metallicity dependent colour, what is actually relevant is the colour ranking of the predicted MS loci with the metallicity; therefore the model predictions would be used in a strictly differential way (only the dependence of the MS colour at ${\rm M_V=6.0}$ on the metallicity is indeed relevant) so partially reducing the problems associated with the zero point of the absolute colour - effective temperature relation.

Let us briefly discuss the impact of the most relevant physical inputs/assumptions used in model computations on the MS colour at ${\rm M_V=6.0}$. This quantity is marginally affected by a change in the adopted radiative opacity table - when one of the updates provided in this last decade is used, and the mixing length parameter is calibrated by computing the appropriate SSM; on the other hand the use of different EOSs can have a large impact of the ${\rm T_{eff}}$ scale of MS stars: when passing from an old EOS to the update FreeEOS one (after recalibrating ${\rm \alpha_{MLT}}$) one can obtain a
${\rm \Delta(B-V)\approx0.05}$ at ${\rm M_V=6.0}$ between the different models.

Actually the location of the fainter portion of the MS locus - being populated by low- and VLM stars with a vanishing extension of the superadiabatic layers - is not significantly affected by a change in the mixing length value: as an order of magnitude the MS colour at  ${\rm M_V=6.0}$  changes (in the optical colours) by $\sim0.01$~mag when changing the mixing length by $\sim0.2$ around the solar calibrated value.

When considering the impact of different heavy element mixtures it is important to consider the case of scaled solar and $\alpha-$enhanced ones: right panel of fig.~7 shows that the MS loci corresponding to two different heavy elements distribution {\it but} the same iron content run almost parallel in the magnitude interval relevant for the application of the MS fitting technique. This occurrence is better shown in fig.~8 that displays
the run of ${\rm (B-V)}$ and the Str\"omgren colour ${\rm (b-y)}$ at ${\rm M_V=6}$  as a function of [Fe/H] (or [M/H]) for scaled solar and $\alpha$-enhanced models: the two sequences are essentially parallel. This implies that the colour shifts are unaffected by the choice of the heavy element mixture, at least in the displayed metallicity interval. 
The ${\rm (b-y)}$ colour appears better suited than ${\rm (B-V)}$, because ${\rm (b-y)}$ is much less sensitive to metallicity, and the size of the colour shifts is minimized.

However, it is important to notice the zero point offset between the scaled solar and $\alpha$-enhanced theoretical 
${\rm (B-V)_{M_V=6}}$  and ${\rm (b-y)_{M_V=6}}$  versus [M/H] relationship. The offset means 
(using again theory just in a differential way) that there would be a systematic colour 
difference between a template MS made of $\alpha$-enhanced stars, and a target scaled solar population at the 
same [M/H]. This offset is reduced at the level of $\sim$0.01~mag in ${\rm (B-V)}$, if the 
comparison between target and template sequence is performed at constant [Fe/H] rather than constant [M/H].

\begin{figure}
\begin{centering}
\vskip -0.5cm
\includegraphics[width=5cm]{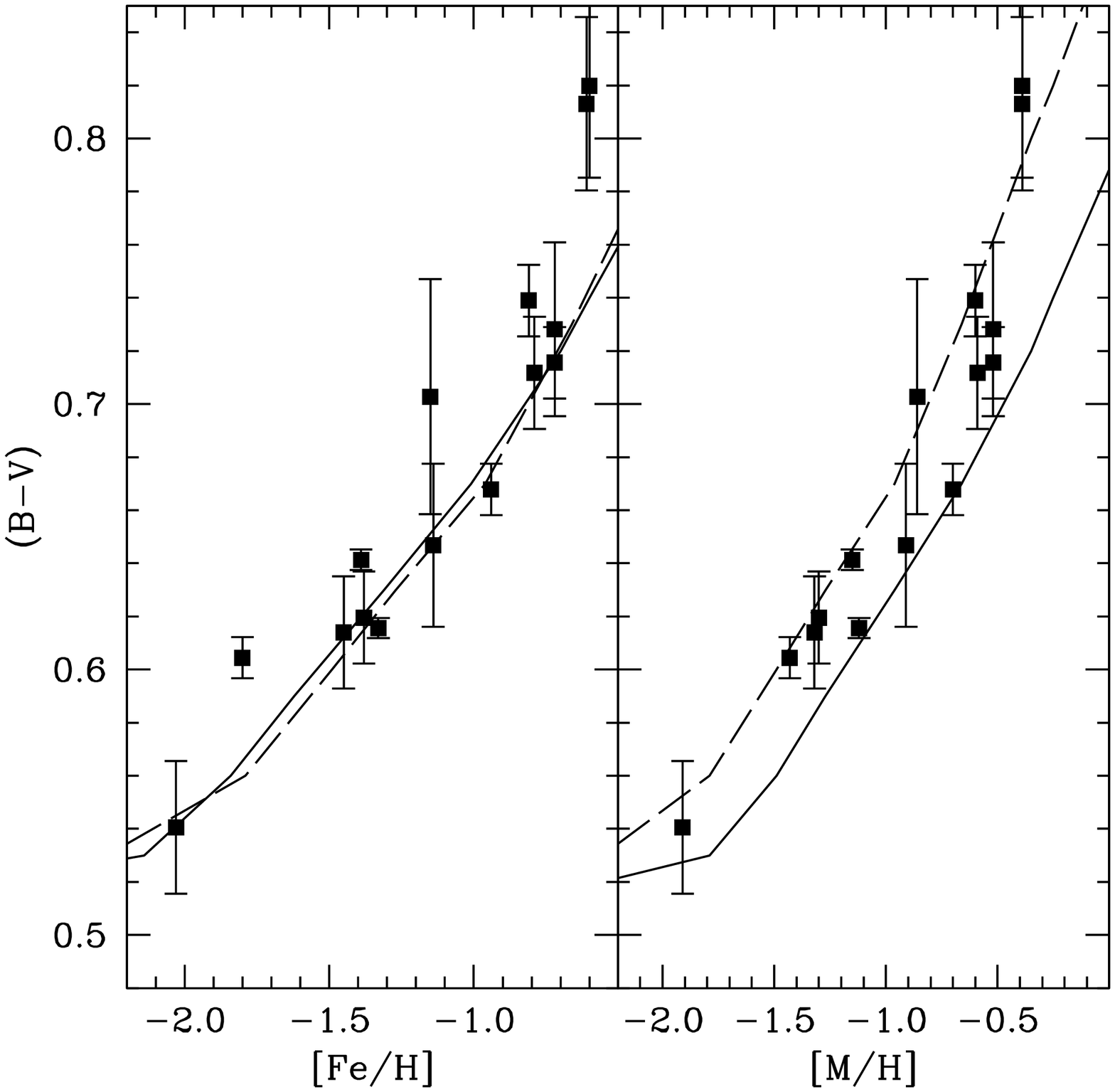}
\qquad
\includegraphics[width=5cm]{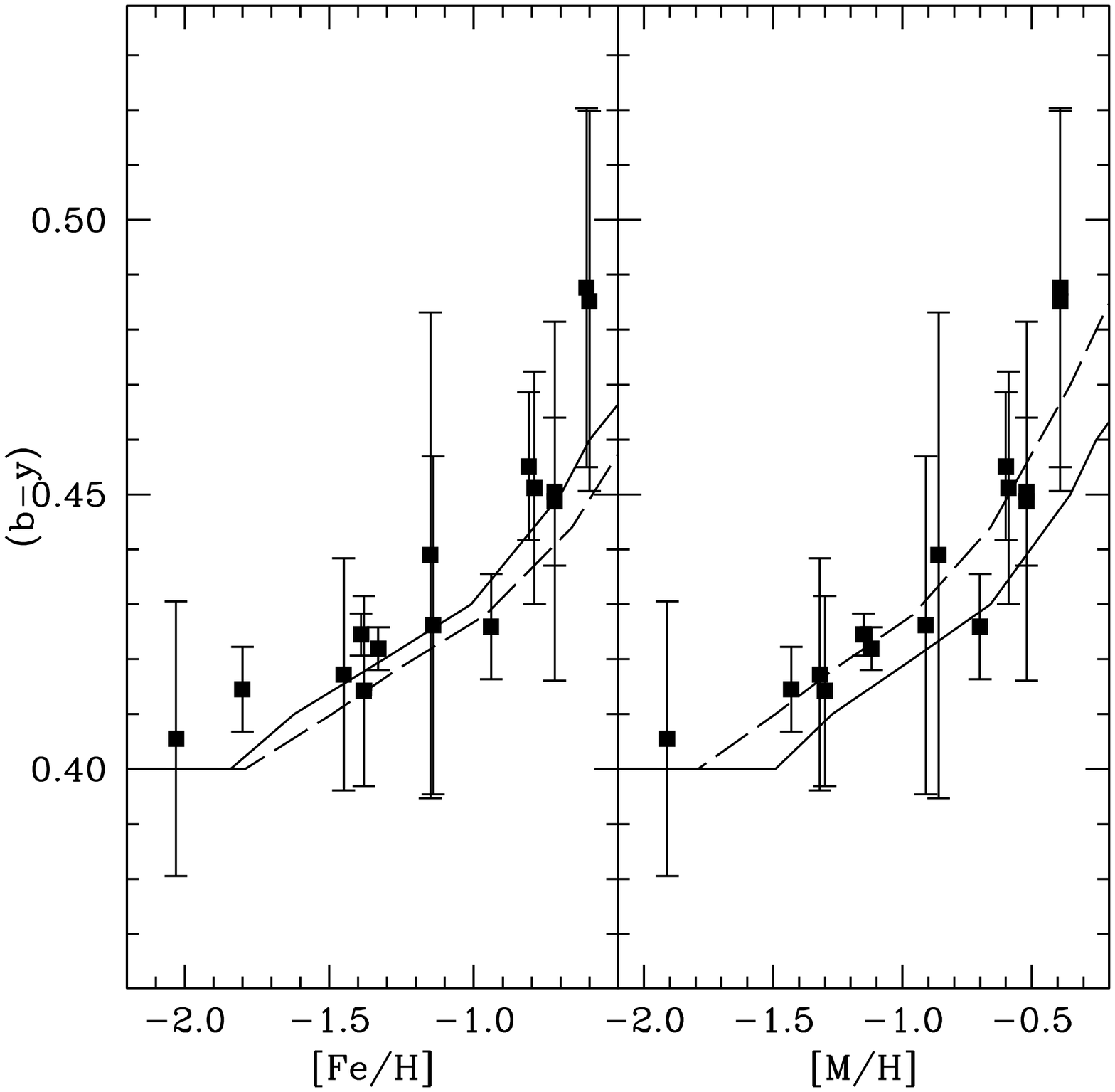}
\vskip -0.5cm
\caption{Theoretical relationship between the ${\rm (B-V)}$ and the Str\"omgren ${\rm (b-y)}$ colour at ${\rm M_V=6}$ and [Fe/H] or [M/H], from the BaSTI $\alpha$-enhanced ([$\alpha$/Fe=0.4], solid lines) and scaled solar (dashed lines) isochrones. Filled squares with errors correspond to empirical data for field subdwarfs.}
\end{centering}
\label{MSfitt3}
\end{figure}

These (or similar) colour-[M/H] relations are used only differentially for calculating  colour shifts to apply to individual subdwarfs. As seen empirically  for stars around solar metallicities, the effect of [Fe/H] (and [M/H]) on colours depends on the selected photometric filter combination. 

An important physical process to consider in terms of its possible impact on MS-fitting distances is 
atomic diffusion. The occurrence of diffusion in the Sun has been amply demonstrated
by helioseismic studies, and it seems that neither turbulence nor other hydrodynamical mixing processes 
substantially reduce the full efficiency of element diffusion in the Sun, otherwise the helioseismic constraints could not be satisfied.
Spectroscopical observations of GGC stars show, however, that atomic diffusion from the convective 
envelope is partially or totally inhibited (see previous discussion on this issue) but, of course, there are no indications whether it is fully efficient in the star interiors.

The effect of fully efficient diffusion on MS-fitting distances to GGCs has been investigated in detail by Salaris et al.~(2000). The crucial point to consider is that  GC spectroscopical metallicity scales are determined on RGB stars, whose surface [Fe/H] is basically unaffected by diffusion (even when fully efficient due to the presence of an extended convective envelope) and is roughly equal to the initial value on the Zero Age MS. However, if the surface metallicity of cluster MS stars is affected by diffusion, it will be typically lower than RGB estimates. As a consequence, when fitting the local template to the MS of a cluster with the same observed
[Fe/H] (measured on the cluster RGB) one is  introducing a potential source of bias in the derived distances. 
Additionally, the colour of the template subdwarfs at the observed spectroscopic [Fe/H] will depend on their (usually unknown) age, for older ages cause a larger decrease of the surface [Fe/H], hence a larger initial Fe content is required to reproduce the observed value of [Fe/H]. This larger initial [Fe/H] affects the present subdwarf  colours and, in turn, the predicted ${\rm \Delta (B-V)_{\rm MS}/\Delta}$ [Fe/H] relationship. These issues can produce an uncertainty on the final distance modulus determination also of the order of
$\approx0.04-0.06$~mag (Salaris et al.~2000).

A final, important issue to be considered is the impact  of multiple stellar populations\footnote{In this last decade, high quality spectroscopic and photometric observations have clearly shown that GGCs do not host a simple stellar populations, but multiple stellar populations characterised by peculiar chemical patterns. In particular - with huge differences from a GC to an other one - there is a plain evidence that GGCs can host a (relevant) fraction of He enhanced stars.} in GGCs, with varying degrees of light element (i.e. C, N, O, Na, etc.) anticorrelation and the -- possibly -- associated increased He abundance, as observed in individual GCs (see Gratton et al.~2012 for a review), but  largely absent in the Halo field subdwarf population.
This issue has been for the first time addressed in the theoretical analysis by Sbordone et al.~(2011). These authors show that the effect of CNONa anticorrelations overimposed to a standard $\alpha$-enhanced mixture at fixed [Fe/H] is negligible in the typical magnitude range employed for MS-fitting distance determination, for photometric filters ranging from U to I and the full set of 
Str\"omgren filters. An enhancement of He compared to the template MS is however able to shift the MS to the blue -- at fixed metal content 
and magnitude -- because of the hotter ${\rm T_{eff}}$ scale (see fig.~13) of the parent stars (bolometric corrections are 
unchanged), causing systematic overestimates of the derived MS-fitting distances. 
Let us consider for example the case of an hypothetical cluster with no obvious multimodality in He content, 
but with a spread of initial He abundances.  Depending on the size of the spread and the number of stars affected, the mean locus 
of the cluster lower MS will be shifted in colour compared to a template sequence built from stars with a uniform 
initial He abundance, and its distance overestimated.

\subsection{The MS turn off}
\label{msto}

The brightness of the bluest point along the MS, the so-called {\sl turn off} (TO),
is the most important clock marking the age of star clusters. The relevance of using this observational feature for the
age dating of star clusters, and more in general, resolved stellar populations has been investigated by many authors
(Cassisi \& Salaris~2013, references therein and other contributions in this volume); therefore we wish to very briefly summarise the main uncertainties affecting the theoretical calibration of the {\sl age - TO brightness} relation.

Let us start remembering that the main \rq{ingredients}\rq\ in stellar models modelling  which affect the age - TO luminosity calibration are the following:

\begin{itemize}

\item{EOS $\longrightarrow$ luminosity, effective temperature}

\item{Radiative opacity $\longrightarrow$ luminosity, effective temperature}

\item{Nuclear reaction rates $\longrightarrow$ luminosity}

\item{Superadiabatic convection $\longrightarrow$ effective temperature}

\item{Chemical abundances $\longrightarrow$ luminosity, effective temperature}

\item{Atomic diffusion $\longrightarrow$ luminosity, effective temperature}

\item{Treatment of the boundary conditions $\longrightarrow$ effective temperature}

\end{itemize}

For each ingredient, the observational property of a TO structure which is
affected by a change of the corresponding ingredient, is also listed. Therefore, it appears evident that some
\lq{inputs}\rq\ affect directly the age - luminosity relation because they modify the bolometric
magnitude of the TO for a fixed age; some other inputs really can modify also (or only) the effective
temperature of the TO models, so they affect the age - luminosity relation indirectly through the
change induced in the bolometric correction adopted for transferring the theoretical predictions
from the H-R diagram to the various observational planes.
In the following we discuss the effect on the calibration of the age - luminosity relation of current uncertainties in some of these inputs .

{\sl Nuclear reaction rates:}
the reliability of theoretical predictions about evolutionary lifetimes
critically depends on the accuracy of the nuclear reaction rates since nuclear burning provides the
bulk of the stellar luminosity during the main evolutionary phases. In these last years, a
large effort has been devoted to increase the measurement accuracy at energies as close as
possible to the Gamow peak, i.e. at the energies at which the nuclear reactions occur in the
stars. 

The effect on the age - luminosity calibration of current uncertainties on the rates of the
nuclear reactions involved in the {\sl p-p} chain is negligible (lower than $\sim2$\%)
 (Chaboyer~1995, Chaboyer et al.~1998). This is due to the fact
 that the nuclear processes involved in the {\sl p-p} chain are really well understood so
the associated uncertainty is quite small.

However, near the end of core H-burning stage, due to the lack of H, the energy supplied by the
H-burning becomes insufficient and the star reacts contracting its core in order to produce the
requested energy via gravitation. As a consequence, both the central temperature and density
increase and, when the temperature attains a value of the order of $\sim15\times10^6$~K, the
H-burning process is really governed by the more efficient {\sl CNO} cycle, whose efficiency is
critically depending on the reaction rate for the nuclear process ${\rm ^{14}N(p,\gamma)^{15}O}$,
the slowest reaction involved in the {\sl CNO} cycle. 

Until few years ago, the rate for this reaction was uncertain, at
least, at the level of a factor of 5. In fact, all available laboratory measurements were
performed at energies well above the range of interest for astrophysical purpose and, therefore,
a crude, and unsafe, extrapolation was required (Angulo et al. 1999, hereinafter NACRE).
Recently the LUNA experiment (Formicola et al. 2003, Marta et al.~2008) has significantly improved
the low energy measurements of this reaction rate, obtaining an estimate which is about a factor
of 2 lower than previous determinations.  The impact of this new rate on the age - luminosity relation is the following: for a fixed TO brightness the new calibration predicts systematically older cluster ages, being the difference
with respect the \lq{old}\rq\ calibration of the order of $\sim0.9$Gyr on average (Imbriani et al. 2004, Weiss et al. 2005, Pietrinferni et al.~2010).

{\sl Diffusive processes:} helioseismological constraints have shown that atomic diffusion must be at work in the Sun. So, it is immediate to assume that this physical
process is also efficient in more metal-poor stars like those currently evolving in GGCs.

This notwithstanding, the evaluation of the atomic diffusion coefficient is not an easy task as a
consequence of the complex physics one has to manage when analizing the various diffusive mechanisms, and
moreover the range of efficiency allowed by helioseismology is still relatively large. For the Sun,
Fiorentini et al. (1999) estimated an uncertainty of about 30\% in the atomic diffusion coefficient.
Perhaps the uncertainty is also larger for metal-poor stars due to the lack of any asteroseismological
constraint. So, it is not unrealistic to estimate an uncertainty of about a factor of two
in the diffusion efficiency. 

From an evolutionary point of view, the larger the atomic diffusion
efficiency, the lower the cluster age estimate is (Castellani et al. 1997). 
As rule of thumb, diffusion reduces the age of old stellar systems by $\sim1$~Gyr, being the effect slightly larger (lower) for the most metal-poor(-rich) stellar populations (due to the different size of the convective envelope, see Salaris \& Cassisi~2005 for more details).

The impact of the uncertainty in the atomic diffusion coefficients on stellar models has been
extensively investigated by Castellani \& Degl'Innocenti (1999):  for ages of the order of 10~Gyr the uncertainty in the efficiency of the microscopic diffusion moves, for each given age, the TO magnitude over a range of $\sim0.16$~mag, affecting the calibration of the age - TO luminosity relation by $?0.7/+0.5$~Gyr.

\begin{figure}
\begin{centering}
\vskip -2.cm
\includegraphics[width=8cm]{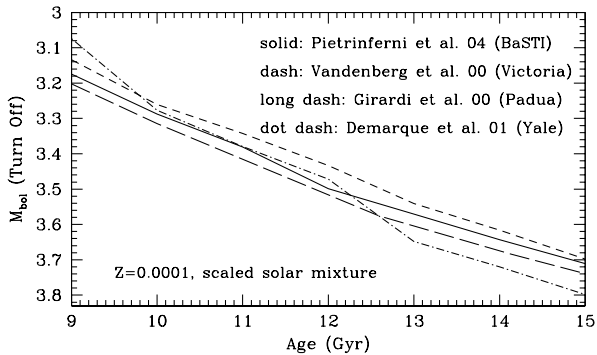}
\vskip -2.cm
\caption{The age - Turn off brightness relation for a given metallicity, as provided by some, widely used, stellar models libraries.}
\end{centering}
\label{tocomp}
\end{figure}

The spectroscopical measurements of iron content in MS GGCs stars as well as the Li and Be abundances in the solar atmosphere (mentioned in section~\ref{ingredient}) support the computations of MS models accounting simultaneously for diffusion, radiative levitation and some sort of extra-mixing, as made by Richard et al. (2002) and Vandenberg et al. (2002): these models are able to reconcile the various observational evidence, but with the need of accounting of some {\sl ad hoc} amount of extra-mixing, whose physical origin is still unclear.

On the other hand, the age - luminosity calibration is not significantly affected by
the inclusion of radiative levitation in the computations: models accounting for both
diffusion and radiative levitation lead to a reduction of the order of 10\% in the GGC age at a given TO
brightness, i.e. more or less the same reduction which is obtained when accounting only for a (standard)
efficiency of microscopic diffusion.

In order to have an idea of the level of confidence in this important age indicator, we show in fig.~9 the TO brightness - age
calibrations provided by some updated sets of evolutionary models: all the theoretical predictions, but the one provided by the Yale group, are in fair agreement. In fact, at a given TO brightness, the  difference in the estimated age is of the order of 0.6-0.8~Gyr or lower, and this difference can be almost completely explained by accounting for the different choices about the initial He content and the physical inputs.

\section{The Red Giant Branch}
\label{rgb}

The RGB is one of the most prominent and well populated features in the CMD of  stellar populations older than about $1.5 - 2$ Gyr.  This 
evolutionary stage is quite important for many reasons as:

\begin{itemize}

\item since RGB stars are cool, reach high luminosities during their evolution, and their evolutionary 
timescales are relatively long, they provide a major contribution to the integrated bolometric magnitude and to integrated colours and spectra at wavelengths larger than about 900 nm of old  distant, unresolved stellar populations (Renzini \& Fusi-Pecci~1988).   
A correct theoretical prediction of the spectral properties of RGB stars is thus of paramount importance for interpreting observations of distant star clusters and galaxies  using population synthesis methods, but also for determining the age of resolved stellar systems by 
means of isochrone fitting techniques;

\item being both the RGB location and slope in the CMD very sensitive to the metallicity, they are widely used as metallicity indicators;
  
\item the I-Cousin band brightness of the tip of the RGB (TRGB) provides a  robust standard candle, largely independent of the stellar age and initial chemical composition, which can allow to obtain  reliable distances out to about 10~Mpc using HST observations (Tammann \& Reindl~2013);

 \item theoretical predictions about the structural properties of RGB stars at the TRGB play a fundamental role in determining the main evolutionary properties of their progeny: core He-burning stars during the Horizontal Branch (HB) evolutionary phase. In particular,  HB luminosities (like the TRGB ones) are mostly determined by the value of the electron degenerate  He-core mass (${\rm M_{core}^{He}}$) at the end of the RGB evolution;
   
\item predicted evolutionary timescales along the RGB phase are an important ingredient in the determination of the initial He abundance 
of GGC stars through the R parameter (Iben~1968a); 

\item an accurate modelling of the mixing mechanisms efficient in RGB stars, such as thermohaline mixing (Charbonnel \& Zahn~2007) is mandatory to correctly  interpret spectroscopic observations of their surface chemical  abundance patterns.    
 
\end{itemize}
     
A detailed discussion on the structural and evolutionary properties of RGB stars, as well as of the level of agreement between theory and observations,  can be found in Salaris \& Cassisi (2005) and Cassisi \& Salaris (2013) and references therein. In the following,
we will briefly review the main observational properties of the RGB such as its ${\rm T_{eff}}$ scale, the
bump of the luminosity function (LF) and the brightness of the Tip; discussing the main sources of uncertainty in the theoretical predictions, and showing some comparison with updated observational data.

\subsection{The effective temperature scale of RGB stars}

The main parameters affecting the ${\rm T_{eff}}$ scale of RGB stars are: EOS, the low-temperature opacity, the efficiency of superadiabatic convection, the outer boundary conditions and the chemical abundances.

{\it EOS:} one of the most widely adopted EOS is the OPAL (Rogers \& Nayfonov~2002) one, whose range of validity does not cover the electron degenerate cores   
of RGB stars and their cooler, most external layers, below 5000 K.  As a consequence, RGB models computed with the OPAL EOS must employ some other EOS to cover the whole thermal stratification of RGB stars. However, there are some notable exception about this as in the case of the models computed by Vandenberg et al.~(2000), Pietrinferni et al. (2004), Dotter et al.~(2007) and many other ones. 
These updated sets of stellar models rely on the use of the FreeEOS  provided by A. Irwin (2005) which consistently allows the computation of stellar models from the H-burning stage to the He-burning phase.

\begin{figure}
\begin{centering}
\vskip -0.5cm
\includegraphics[width=10cm]{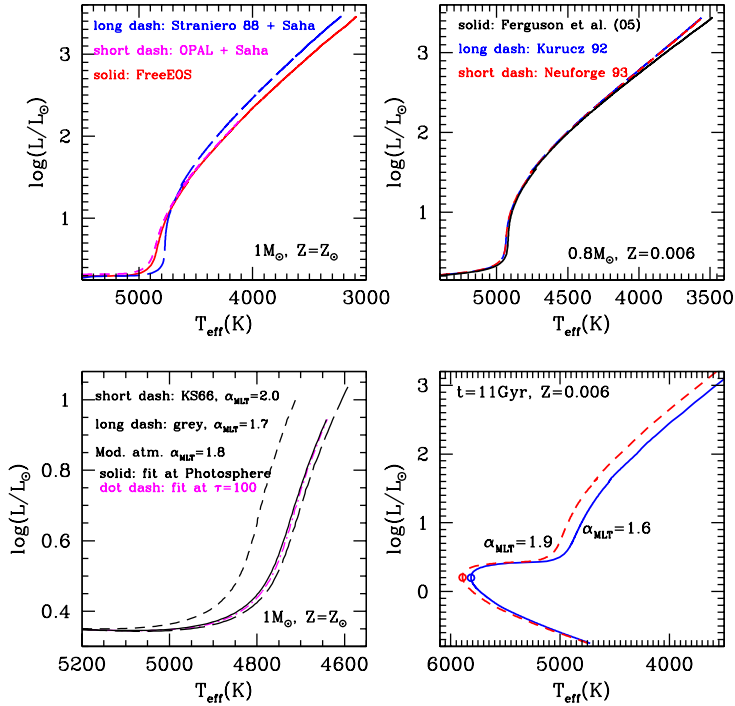}
\vskip -0.5cm
\caption{{\sl Upper left panel:} the RGB portion of the evolutionary track of solar metallicity ${\rm 1M_\odot}$ models computed by adopting different EOSs. 
{\sl Upper right panel:}  the RGB loci of ${\rm 0.8M_\odot}$ stellar model for a metallicity Z=0.006, computed by using different prescriptions about the low-T radiative opacities.
{\sl Lower left panel:}  the sub-giant branch and RGB portion of evolutionary tracks of various SSMs computed with various outer boundary conditions (see labels). For each model it has been also listed the corresponding value of the mixing length parameter. {\sl Lower right panel:} two isochrones based on stellar models computed by using the same physical inputs but different values for ${\rm \alpha_{MLT}}$.}
\end{centering}
\label{rgb}
\end{figure}

The upper left panel of fig.~10 shows the RGB of a ${\rm 1M_\odot}$ stellar structure computed
adopting different assumptions about the EOS: the models based on the OPAL EOS
and the FreeEOS are in very good agreement (this comparison is actually meaningful only for ${\rm T_{eff}}$
larger than about 4500K for the reason discussed before); while there is a significant change in the RGB slope
with respect the model based on the Straniero (1988) EOS supplemented at the lower temperature by a Saha
EOS. On average, there is a difference of about 100K between RGB models based of the two different EOSs.

{\it Low-T opacity:} it has been shown by Salaris et al~(1993) that it is the low-T opacity which mainly determines the   
${\rm T_{\rm eff}}$ location of RGB models, while the high-T one - in particular opacity for temperatures around $10^6$~K - enter in the determination of the mass extension of the convective envelope.
Present generations of stellar models employ mainly the low-T opacity calculations 
by Ferguson et al.~(2005) which are the most  up-to-date computations suitable for stellar modelling, spanning a wide range of chemical compositions.     
The main improvement of this low-T opacity set with respect older evaluations is  in the treatment of molecular absorption and for a proper
treatment of the presence of grains. Although significant improvements are still possible as a consequence of a better
treatment of the different molecular opacity sources, we do not expect dramatic changes in the temperature
regime where the contribution of atoms and molecules dominate, whilst large variations can be foreseen in the regime
(${\rm T<2000}$~K) where grains dominates the interaction between radiation and matter.

Upper right panel of fig.~10 shows a comparison between RGB models computed by adopting the same physical inputs but different low-T opacity evaluations: for effective temperature larger than $\sim$4000~K, the ${\rm T_{eff}}$ scale of the models computed by using the (nowadays obsolete) opacities provided by Neuforge~(1993) and Kurucz~(1992) - providing almost   undistinguishable results - is in very good agreement with that corresponding to the model computed by adopting the Ferguson et al.~(2005) opacities. However, as soon as the RGB ${\rm T_{\rm eff}}$ goes below this limit  (i.e., when the models approach the TRGB and/or their initial metallicity is increased),  the Ferguson et al.~(2005) opacities produce progressively cooler models (differences reaching values of the order of 100 K or more),  due to the increasing effect of the molecules (in particular $\rm H_{2}O$ and TiO) which contribute substantially to the (increase of the) opacity in this temperature range. 

{\it Outer boundary conditions:} the procedure commonly adopted for fixing the boundary conditions at the bottom of the atmospheric layers, relies in the integration of the
atmosphere by using a functional (semi-empirical or theoretical)  (${\rm T(\tau)}$) relation between the temperature and the
optical depth $\tau)$. The effect of using boundary conditions from model atmospheres on RGB stellar models have been 
recently investigated by Vandeberg et al.~(2008).
In lower left panel of fig.~10 it is shown the effects of different ${\rm T(\tau)}$ relations, namely, the Krishna-Swamy~(1966) solar ${\rm T(\tau)}$ relationship and the grey one, on a ${\rm 1M_\odot}$ solar metallicity model computed by using the same physical inputs and a solar calibrated mixing length parameter.  By construction all tracks overlap at the present solar luminosity and effective temperature (actually - although not shown - the agreement is very good also along the whole MS). On the RGB the track based on the KS66 atmosphere are $\sim100$ K hotter than the tracks based on the model atmosphere
stratification (in the specific case the MARCS models have been used - see Vandenberg et al.~2008 for more details) and the Eddington ${\rm T(\tau)}$ relationship. It is also very important to notice that the two tracks obtained matching the MARCS atmospheres to the interior structure at different optical depths: $\tau  = 1$ and 100, respectively, do overlap from the Zero Age MS to the RGB. This occurrence is also valid at lower metallicities, and is a proof that the stellar model ${\rm T_{eff}}$ scale does not depend on the chosen fitting point until a value of $\tau$ larger or equal to unity is chosen.
When comparing more metal-poor, low-mass stellar models Ð based on solar-calibrated ${\rm \alpha_{MLT}}$ Ð, the difference in the ${\rm T_{eff}}$ scale between models based on both types of grey atmospheres and detailed non-grey models appears negligible. It is worth noting that the comparison between theory and observations for the effective temperature scale of approximately solar metallicity RGB stars both in the field and in star clusters, seems to support the ${\rm T_{eff}}$ scale of stellar models based on the KS66 ${\rm T(\tau)}$ relation.

Before closing this discussion, we wish to remark that, even if the use of boundary conditions provided by model atmospheres is, in principle, more rigorous, one has also to bear in mind that the convection treatment in the  adopted model atmospheres (Montalban et al.~2001) is usually not the same as in the underlying  stellar models (i.e., a different mixing length formalism and a different value for the scale height of the convective motion are used).
 
{\it The treatment of superadiabatic convection:}  as already discussed the value of ${\rm \alpha_{\rm MLT}}$ is usually calibrated by reproducing the solar radius, and this  solar-calibrated value is then used for stellar models of various masses and along different evolutionary phases, including the RGB one. 
This procedure guarantees that models always predict correctly the ${\rm T_{\rm eff}}$  of, at least, solar type stars. However, 
the RGB location is much more sensitive to the value of $\alpha_{\rm MLT}$ than the MS - as shown in the lower right panel of fig.~10. This is due to the evidence that along the RGB the extension (in radius) of the superadiabatic layers - as a consequence of the much more expanded configuration achieved by the star - is quite larger in comparison with the MS configuration.
Therefore, it is important to verify that a solar calibrated ${\rm \alpha_{\rm MLT}}$ is always suitable also for RGB stars of   
various metallicities.   
  
The strong dependence of the RGB model ${\rm T_{eff}}$ scale on the treatment of superadiabatic convection offers an independent method for calibrating ${\rm \alpha_{\rm MLT}}$: one can compare empirical estimates of ${\rm T_{eff}}$ Ð at a fixed absolute magnitude Ð for the RGB of GGCs with theoretical models of the appropriate chemical composition and various assumptions on ${\rm \alpha_{\rm MLT}}$. Left panel of fig.~11 shows a comparison between the empirical RGB ${\rm T_{eff}}$ for a sample of GGCs and model predictions based on a solar calibrated ${\rm \alpha_{\rm MLT}}$ and the two different flavours of the MLT (Salaris \& Cassisi~2008). One can easily note that: i) a solar calibrated ${\rm \alpha_{\rm MLT}}$ provides also a satisfactory match to the effective temperatures of RGB metal-poor stars; and ii) the effective temperature scale of RGB stellar models obtained with the ML2 flavour of the MLT and the solar calibrated value of ${\rm \alpha_{MLT}}$ are  $\sim50$ K hotter than BV58 results, but both model ${\rm T_{eff}}$ scales appear consistent with empirical results within current uncertainties.
   
It is important to emphasize that there is in principle no reason why ${\rm \alpha_{\rm MLT}}$ should be kept constant when considering stars of different masses and/or chemical composition and/or at different evolutionary stages; even within the same star ${\rm \alpha_{\rm MLT}}$ might in principle vary from layer to layer.

One has also to keep in mind that observational uncertainties can affect the MLT calibration based on the empirical RGB ${\rm T_{eff}}$ scale but the effect is usually small: for instance, although to determine the RGB ${\rm T_{eff}}$ at a fixed absolute magnitude of a given GC, one has to rely on GC distance estimates, because the RGB is roughly vertical, an uncertainty of $\sim\pm0.2$ mag in the individual cluster distance modulus does not affect at all the MLT calibration. The same outcome applies for the presence of random uncertainties on the cluster reddening and metallicity.
To give an estimate of the sensitivity of the calibrated ${\rm \alpha_{\rm MLT}}$ value to systematic errors on the temperature scale, chemical composition, and adopted distances, one can note that systematic changes of the empirical RGB ${\rm T_{eff}}$ by $\sim70$ K, or cluster [M/H] by 0.2 dex, or adopted cluster distance moduli by 0.25 mag would cause a change in the calibrated ${\rm \alpha_{\rm MLT}}$ of about $\sim0.1$.

\begin{figure}
\begin{centering}
\vskip -0.5cm
\includegraphics[width=5.6cm]{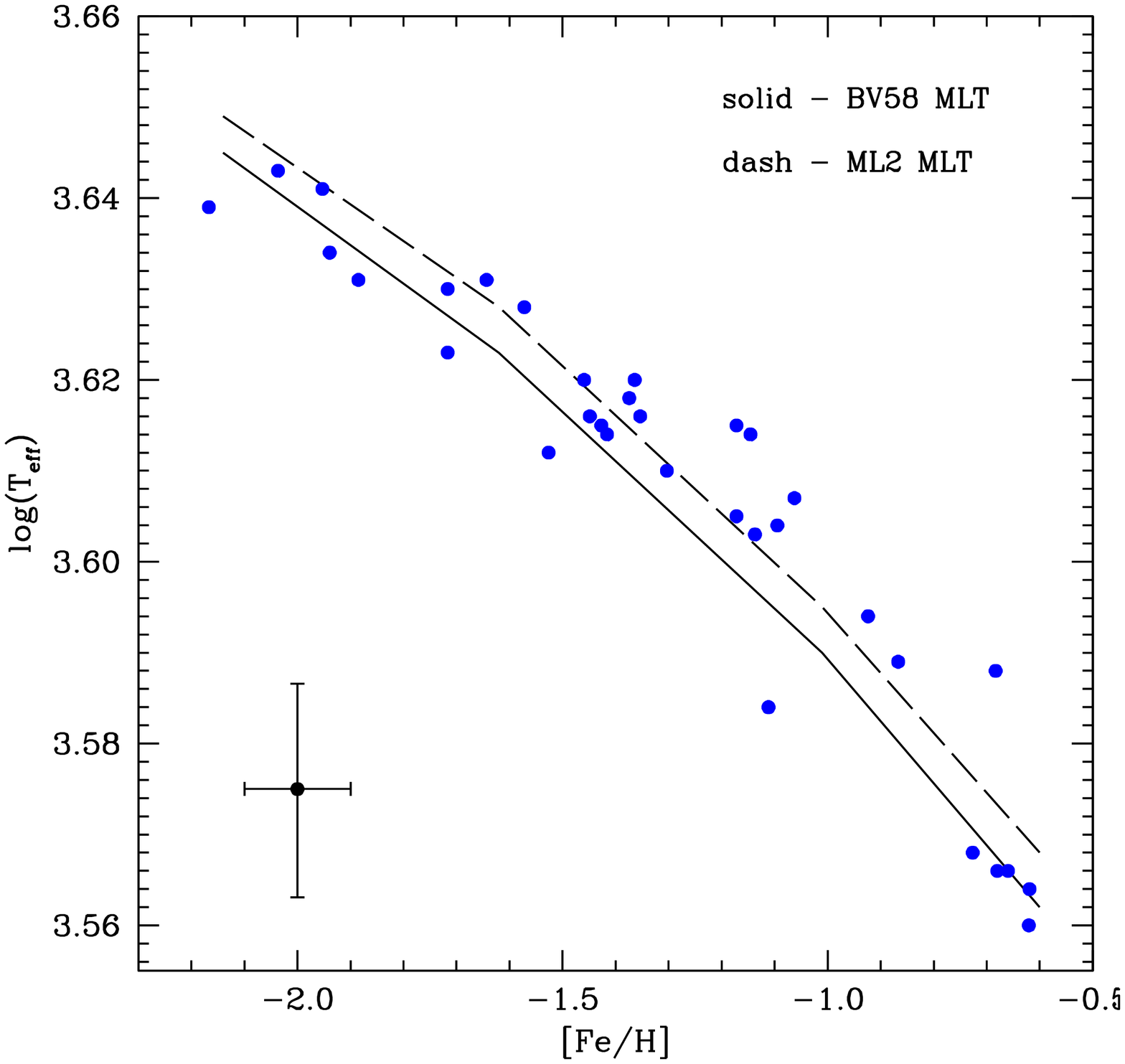}
\qquad
\includegraphics[width=5.7cm]{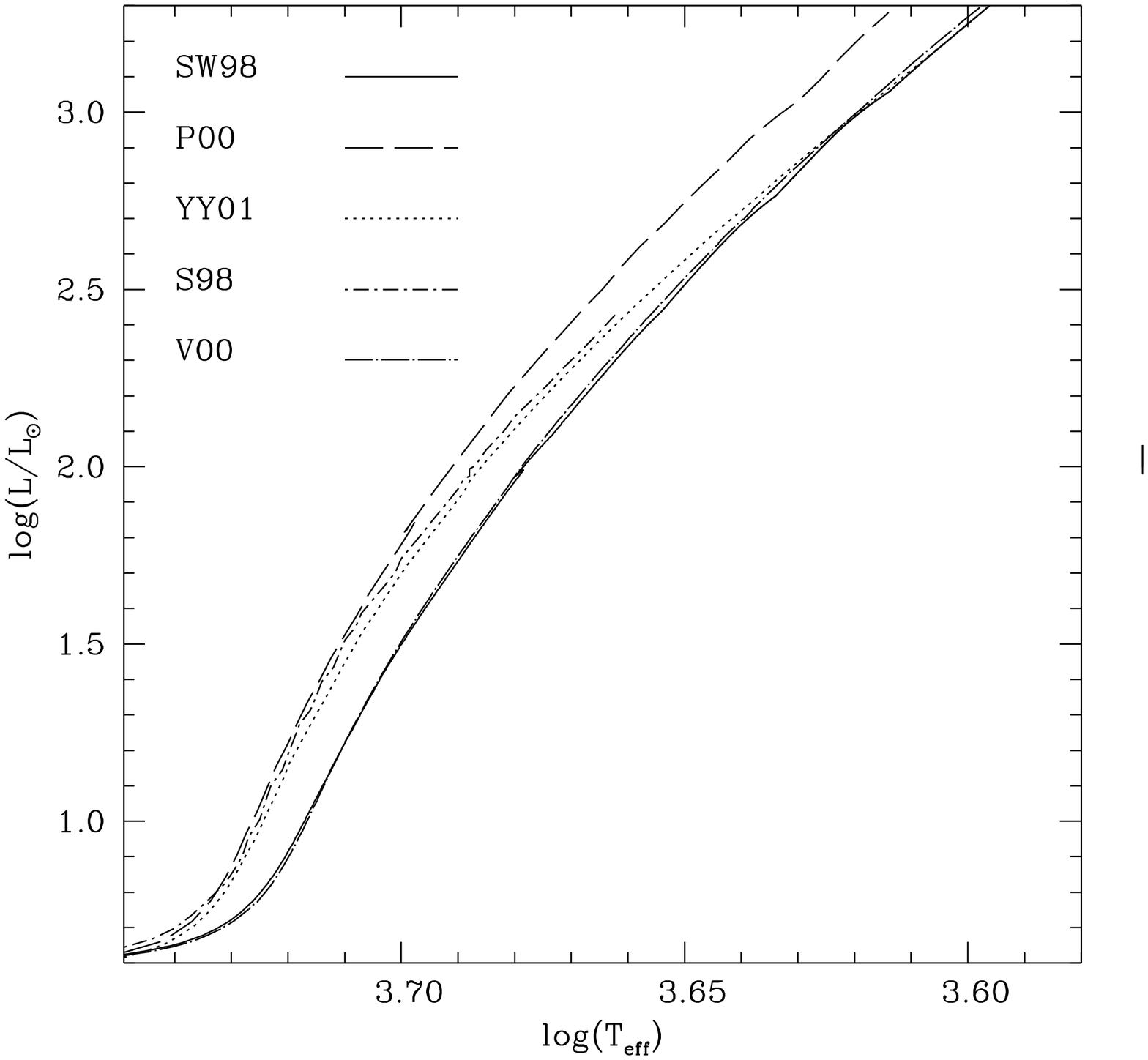}
\vskip -0.5cm
\caption{{\sl Left panel:} empirical estimates of the average RGB ${\rm T_{eff}}$ at ${\rm  M_{bol} =-3}$, for a sample of GGCs as a function of the metallicity. The two lines correspond to theoretical predictions obtained from RGB stellar models computed by using alternatively the solar calibrated BV58 and ML2 flavours of the MLT. The point with the error bar shows the typical observational uncertainty.
{\sl Right panel:} the RGB portion of isochrones computed with a solar-calibrated MLT, for a scaled solar metal mixture and  an age of 12~Gyr, as provided from different authors: Girardi et al. (2000, P00), Yonsei-Yale models (Yi et al.~2001, YY01), VandenBerg et al. (2000, V00), Salaris \& Weiss (1998, SW98), and the FST models by Silvestri et al. (1998, S98).
}
\end{centering}
\label{rgb2}
\end{figure}
  
This notwithstanding, a source of concern about an {\sl a priori} assumption of a solar $\alpha_{\rm MLT}$   
for RGB computations comes from the fact that models from various authors, all using a suitably     
calibrated solar value of $\alpha_{\rm MLT}$, do not show the same RGB temperatures.    
This means that -- for a fixed RGB temperature scale --  the calibration of $\alpha_{\rm MLT}$ on the empirical RGB ${\rm T_{eff}}$ values would not provide always the solar value.   

Figure~11 displays also isochrones produced by independent groups, all computed with the same initial chemical composition, same opacities, and the appropriate solar calibrated values of $\alpha_{\rm MLT}$: while the Vandenberg et al. (2000) and Salaris \& Weiss (1998) models are identical, the Padua ones (Girardi et al. 2000) are systematically hotter 
by $\sim$200 K, while the $Y^2$ ones (Yi et al.~2001) have a different shape. This comparison shows clearly that if one set of MLT solar calibrated RGBs can reproduce a set of empirical RGB temperatures, the others cannot, and therefore in some case a solar calibrated $\alpha_{\rm MLT}$  value may not be adequate.  
The reason for these discrepancies can be due to some  difference in the input physics, like the EOS and/or the boundary conditions, which is not compensated by the solar recalibration of $\alpha_{\rm MLT}$ (see data shown in the lower left panel of fig.~10).   
  
This occurrence clearly points out the fact that one cannot expect the same RGB ${\rm T_{eff}}$ from solar calibrated models  
not employing exactly the same input physics. The obvious conclusion is that it is always necessary to compare RGB models with observations to ensure the proper calibration of ${\rm \alpha_{MLT}}$ for RGB stars.

\begin{figure}
\begin{centering}
\vskip -0.5cm
\includegraphics[width=8.cm]{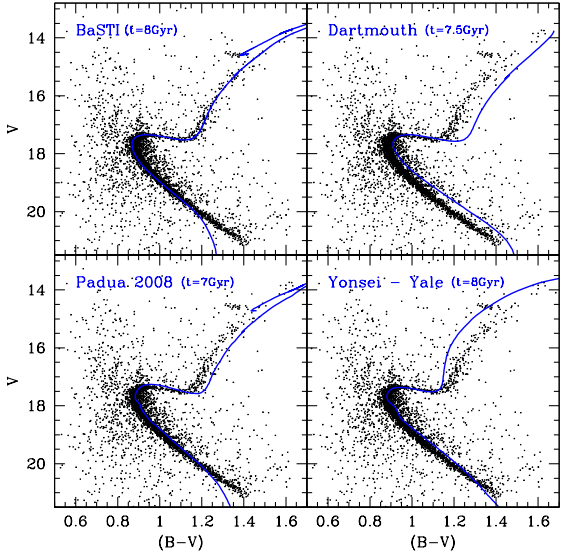}
\vskip -0.5cm
\caption{Comparison between the CMD of the old, open cluster NGC~6791 (Stetson et al.~2003) and isochrones from
independent stellar model libraries for a metallicity appropriate to the cluster (${\rm [Fe/H]\approx+0.4}$). In each case, the adopted distance modulus and reddening have been allowed to vary within the estimated $1\sigma$ uncertainty.}
\end{centering}
\label{ngc6791}
\end{figure}

Before closing this section we wish to note that the fact that a set of stellar models reproduce fairly well the ${\rm T_{eff}}$ scale of cluster RGB stars is not a warranty that the same holds also when comparing the evolutionary predictions 
with data in the various CMDs. In fact, in order to compare model predictions - such as isochrones - with observations, the theoretical framework has to be transferred from the H-R diagram to the chosen observational plane by using a suitable colour - ${\rm T_{eff}}$ relations, and one has to check the capability of the predicted colours and magnitudes to fit the empirical sequences in the observational CMD. The reliability of this check is usually strongly reduced by the fact that the large, current uncertainties on both cluster distance, reddening and metallicity, allow to tune the value of these quantities in order to obtain a satisfactory fit. However, if the distance modulus, reddening and metallicity are known with high accuracy as in the case of the
old (${\rm \sim8}$~Gyr), open cluster NGC~6791 (Grundahl et al.~2008), the degrees of freedom in the fitting procedure are hugely reduced and this allows a meaningful check of the capability of the theoretical framework (combination of stellar evolutionary models and colour - ${\rm T_{eff}}$ relations) to reproduce the observations.

An example of this kind of experiment has been performed in fig.~12, showing the comparison between the same photometric dataset for NGC~6791 and various stellar model libraries obtained by allowing the cluster reddening, distance and metallicity to vary within the estimated $1\sigma$ uncertainty. From data shown in this figure, one can note that same models retrieved from publicly available stellar model libraries are able to match the location of the various sequences in the CMD whereas other ones do not.

{\it The chemical composition:}  the heavy element abundance is one of the parameters which most affects the RGB morphology: any increase of $Z$ produces a larger envelope opacity and, in turn, a more extended envelope convection
zone and a cooler RGB. The strong dependence of the RGB effective temperature on the metallicity makes the RGB one of the most important metallicity indicators for stellar systems. 
An important issue is the dependence of the shape and location  of the RGB on the distribution of the metals: different heavy elements have different ionization potentials, and provide different contribution to the envelope opacity. The abundance of low ionization potential elements such as Mg, Si, S, Ca, Ti and Fe strongly influences the RGB effective temperature, through their
direct contribution to the opacity due to the formation of molecules such as TiO which strongly affects the stellar spectra at effective temperatures lower than $\sim5000$~K, and through the electrons released when ionization occurs, which affect the envelope opacity via the formation of the H$^-$ ion -- one of the most important opacity sources in RGB structures (see Pietrinferni et al.~2009, Sbordone et al.~2011, Cassisi et al.~2013, and Vandenberg et al.~2012, for a recent analysis of this topic). As an example, a change of the heavy elements mixture from a scaled solar one to an $\alpha$-element enhanced distribution with the same iron content, produces a larger envelope opacity and the RGB becomes cooler and less steep: the change in the slope being due to the increasing contribution of molecules to the envelope opacity when the stellar ${\rm T_{eff}}$ decreases along the RGB.

\begin{figure}
\begin{centering}
\vskip -0.6cm
\includegraphics[width=7.5cm]{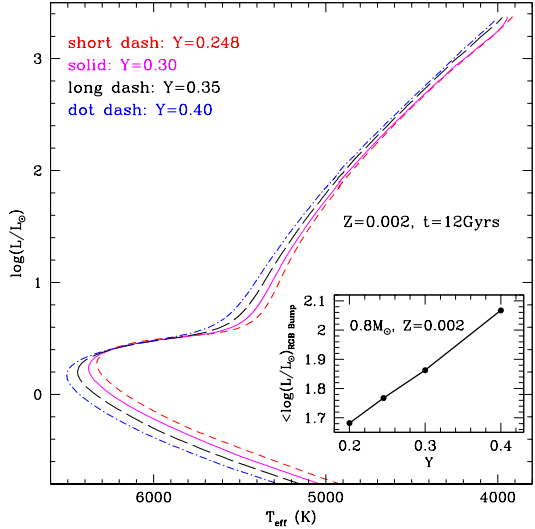}
\qquad
\vskip -0.7cm
\caption{Comparison between 12~Gyr old, Z=0.002, isochrones computed for various assumptions about the initial He abundance. The inset shows the trend of the average RGB bump brightness as a  function of the initial He abundance for a ${\rm 0.8M_\odot}$ model.}
\end{centering}
\label{elio}
\end{figure}

As far as it concerns the helium abundance, this element is one of the most important chemical species in the context of stellar evolution, because any change of its abundance hugely affects the structural and evolutionary properties of stars. In more detail, a change of the He abundance affects low-T, radiative opacities, for an increase of He causes a reduction of the opacity: when increasing the He abundance at fixed metallicity, the H abundance has to decrease, and given that H is a major opacity source via the ${\rm ^-H}$ ion, this causes a global reduction of the opacity. This effect explains why He-enhanced stellar models have hotter ${\rm T_{eff}}$ values in comparison with normal He abundance stars. 

The opacity reduction due to He-enhancement contributes also to make brighter the stars during both the core H-burning and He-burning (for HB stars with a massive enough envelope, see Cassisi \& Salaris~2013 for a detailed discussion on this issue) stages, although the larger contribution to the change in the stellar surface luminosity is due to the change in the mean molecular weight ($\mu$) associated to the He abundance increase. In fact, the H-burning efficiency is strongly dependent on the value of the mean molecular weight: ${\rm L_H\propto\mu^7}$. .

Figure~13 shows different isochrones computed with the same [Fe/H] and age, but various values for the initial He content. There are some interesting features disclosed by this plot: {\sl i)} the He-rich MS runs parallel in the luminosity interval from the MS TO and the location of stars with mass ${\rm \sim0.5M_\odot}$; {\sl ii)} the MS effective temperature is sensitive to the He increase (${\rm \Delta{T_{eff}}}/{\Delta{Y}}\approx 2.3\times10^3 K$); {\sl iii)} the ${\rm T_{eff}}$ values of RGB stellar models are also affected by an He increase, although to a smaller extent than the MS locus. 
One has also to note that, at fixed age, the mass at the MS TO significantly decreases when increasing the initial He content: for a 12Gyr old isochrone it is equal to ${\rm \sim0.81M_\odot}$ for Y=0.245 and ${\rm \sim0.61M_\odot}$ for Y=0.40; this occurrence has very important implications on the morphology of the horizontal branch for He-enhanced stellar populations.
   
\subsection{The Red Giant Branch luminosity function}
\label{bump}

The RGB luminosity function (LF), i.e. the number of stars per brightness bin among the RGB as a function
of the brightness itself, of GGCs is an important tool to test the chemical stratification inside
the stellar envelopes (Renzini \& Fusi Pecci 1988).
The RGB LF is a simple straight line on a magnitude-$\log(N)$ plane, except in a very narrow magnitude interval (see below). This linearity is a direct consequence of the He-core mass - luminosity relation for RGB stars (Cassisi \& Salaris~2013). The slope of this line allows a major test of the evolutionary rate along the RGB, that is virtually independent of the isochrone age or metallicity. Recent comparisons between theoretical LFs and empirical ones for GGCs (e.g., Sandquist et al.~2010) have shown a remarkable good agreement: this represents a plain evidence of the fact that theoretical predictions about the evolutionary rate along RGB are quite reliable.

\begin{figure}
\begin{centering}
\vskip -1.cm
\includegraphics[width=8.cm]{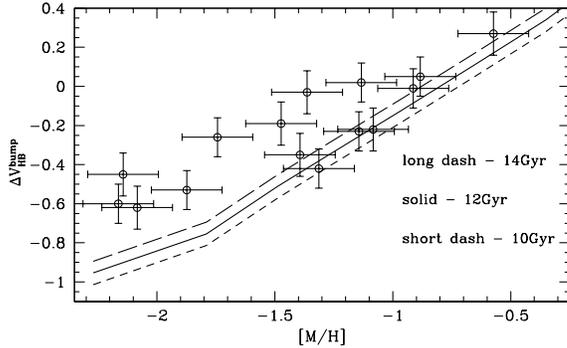}
\qquad
\vskip -3.3cm
\caption{Comparison between empirical measurements of the brightness difference between the RGB bump and the ZAHB and BaSTI theoretical prescriptions - for the labeled assumptions about the cluster age - as a function of the global metallicity.}
\end{centering}
\label{bumphb}
\end{figure}

The most interesting feature of the RGB LF is the occurrence of a local maximum in the luminosity distribution of RGB stars, which appears as a bump in the differential LF, and as a change in the slope of the cumulative LF. This feature is caused by
the sudden increase of H-abundance left over by the surface convection upon reaching its maximum inward extension at
the base of the RGB (\emph{first dredge up}) (Thomas~1967, Iben~1968b). When the advancing H-burning shell encounters this
discontinuity, its efficiency is affected (sudden increase of the available fuel), causing a temporary drop of the surface luminosity. After some time the thermal  equilibrium is restored and the surface luminosity starts to increase again. As a consequence, the stars cross the same luminosity interval three times, and this occurrence shows up as a characteristic peak in the differential LF. Moreover, since the H-profile before and after the discontinuity is different, the rate of advance of the H-burning shell changes when the discontinuity is crossed, thus causing a change in the slope of the cumulative LF.

The brightness of the RGB bump is therefore related to the location of this H-abundance discontinuity, in the sense that the deeper the chemical discontinuity is located, the fainter is the bump luminosity. Therefore, the RGB bump luminosity depends significantly
on the stellar chemical composition: an increase of the global metallicity - at a given age and initial He abundance - causes a decrease
of the bump brightness because, due to the larger opacity, the H-abundance discontinuity is located in deeper layers; when considering an increase of the initial He abundance - being all the other parameters kept fixed - model computations predict: {\sl a)} an increase of the bump brightness (see the inset in fig.~13), and {\sl b)} a smaller luminosity excursion during the RGB bump stage. The first effect is due to the lower envelope opacity of He-rich stars, that causes the discontinuity of the H abundance left over by the \emph{first dredge up} to be located in more external layers. The second effect is caused by the fact that in He-rich stars, the jump of the H abundance at the discontinuity is smaller than in normal He stars. As a consequence the surface stellar luminosity is less affected when the H-burning shell crosses the discontinuity.

Since any physical inputs and/or numerical assumption adopted in the computations which affects the maximum extension of the convective envelope at the \emph{first dredge up}, strongly affects the bump brightness.
Therefore, a comparison between the predicted bump luminosity and the observations allows a direct check of how well theoretical models for RGB stars predict the extension of convective regions in the stellar envelope and, then provide a
plain evidence of the reliability of current evolutionary framework.

However, when comparing theory with observations one needs a preliminary estimate of both the cluster metallicity and
distance. Current uncertainty in the GGC metallicity scale strongly reduces our capability to 
constrain the plausibility of the theoretical framework, and for such reason,
it has became a common procedure to use simultaneously all available metallicity scales (Riello et al. 2003). 
An other critical issue is related to the knowledge of the cluster distance, whose indetermination could strongly hamper the
possibility of a meaningful comparison between theory and observations. In order to overcome this problem, as early suggested  
by Fusi Pecci et al.~(1990), the observed V-magnitude (or filters similar to Johnson V) difference between the RGB bump and the HB at the RR Lyrae instability strip (\dvbump) is usually adopted in order to test the theoretical
predictions for the bump brightness. This quantity presents several advantages from the
observational point of view (Fusi Pecci et al.~1990, Salaris et al.~2002) and it is empirically well-defined
because it does not depend on a previous knowledge of the cluster distance and reddening. However, on the theoretical
side, one should keep in mind that such comparison requires the use of a theoretical prediction about the Horizontal
Branch brightness which is a parameter still affected by some uncertainty (see section~\ref{hestage}). Nevertheless, 
empirical estimates about the \dvbump parameter have been extensively compared with theoretical predictions (Di Cecco et al. 2010 and references therein). Figure~14 shows a comparison between the most recent measurements of this
parameter for a large sample of GGCs and theoretical predictions, and discloses a discrepancy Ð at the level of $\sim0.20$~mag, or possibly larger Ð, for GCs with [M/H] below $\sim-1.5$: the predicted \dvbump being larger than observed. We cannot discriminate whether this is due to too bright theoretical bump luminosities, or under-luminous HB models (or a combination of both effects). Due to the strong dependence of \dvbump parameter on metallicity, a quantitative assessment of the actual discrepancy between theory and observation strongly depends also on the adopted metallicity scale. 

\begin{figure}
\begin{centering}
\vskip -0.8cm
\includegraphics[width=8.cm]{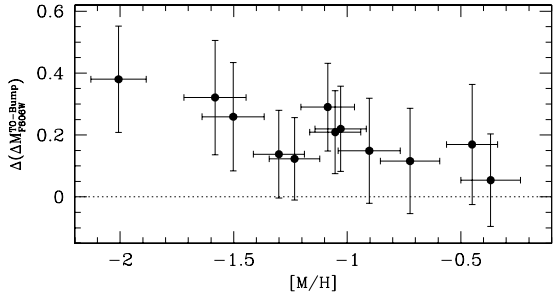}
\qquad
\vskip -0.5cm
\caption{Difference between the values of  ${\rm \Delta{M_{F606W}}^{Bump}_{TO}}$ predicted by models for each individual GC - once its age has been obtained from the absolute TO brightness -, and the measured values, as a function of [M/H] (see text for more details).}
\end{centering}
\label{bumpto}
\end{figure}

A complementary avenue is offered by the magnitude difference between the TO and RGB bump, ${\rm \Delta{V}^{Bump}_{TO} = V_{TO} - V_{bump}}$, which bypasses the HB. Figure~15 shows a comparison between this quantity as measured for a sample of GGCs observed in the F606W filter of the Advanced Camera for Surveys on board of HST and the similar theoretical quantity as estimated once the age of each GC has been accurately derived via the isochrone fitting to the absolute MS TO magnitude obtained using a MS-fitting distance scale (see Cassisi et al. 2011, for more detailed discussion on the adopted method). This plot shows clearly that the expected ${\rm \Delta{M_{F606W}}^{Bump}_{TO}}$ values are systematically larger than observed: the mean difference being of the order of 0.20~mag. Given that the observed TO magnitude is by definition matched by the theoretical isochrones to determine the TO age, this discrepancy implies that the absolute magnitude of the RGB bump in the models is too bright, e.g., RGB models predict a too shallow maximum depth of the envelope at the first dredge up. This explains also the similar discrepancy found for the  \dvbump parameter.

On the basis of the most recent updates in the input physics, it does not appear realistic to modify the maximum depth of the convective envelope at the dredge up by reasonable changes in the adopted physics (i.e., the radiative opacity). The most viable solution of this discrepancy implies some amount of convective overshoot, beyond the Schwarzschild convective boundary, by about 0.25${\rm H_P}$.

We note that the RGB LF bump provides other important constraints besides brightness for checking the accuracy of RGB models. More in detail, both the shape and the location of the bump along the RGB LF can be used for investigating on how \lq{steep}\rq\ is the H-discontinuity left over by envelope convection at the \emph{first dredge
up}. So these features appear, potentially, a useful tool for investigating on the efficiency of non-canonical mixing at the border of the convective envelope (Cassisi et al.~2002) able to partially smooth the chemical discontinuity. In addition, since the evolutionary rate along the RGB is strongly affected by any change in the chemical profile, it is clear that the star counts in the bump region can provide reliable information about the size of the jump in the H profile left over by envelope convection after the \emph{first dredge up} (Bono et al.~2001, Riello et al.~2003).

\subsection{Tip of the Red Giant Branch}

As a consequence of the H-burning occurring in the shell, the mass size of the He core ($M_{core}^{He}$) grows. 
When ${\rm M_{core}^{He}}$ reaches about 0.50 $M_{\odot}$ (the precise value depends weakly on the total mass of the star for structures less massive than ${\rm \sim1.2M_\odot}$, i.e. older than about 4-5~Gyr, being more sensitive    
to the initial chemical composition), He-ignition occurs in the   
electron degenerate core. This process is the so called He-flash, that terminates the RGB phase   
by removing the electron degeneracy in the core, and driving the star onto its Zero Age Horizontal Branch (ZAHB) location, that   
marks the start of quiescent central He-burning. The brightest point along the RGB, that marks the He ignition through the He flash is the so-called Tip of the RGB (TRGB).

The observational and evolutionary properties of RGB stars at the TRGB play a pivotal role in current stellar astrophysical
research. The reasons are manifold: i) the mass size of the He core at the He flash fixes not only the TRGB brightness but
also the luminosity of the Horizontal Branch, ii) the TRGB brightness is one of the most important primary distance indicators, iii) the TRGB luminosity can be used as a benchmark for fundamental physics, such as an example the value of the putative neutrino magnetic dipole moment (e.g.,  Viaux et al.~2013).

The reasons which make the TRGB brightness a quite suitable standard candle are the following: the TRGB
luminosity that is a strong function of the He core mass at the He flash, is weakly dependent on the stellar mass, and therefore on the cluster age over a wide age interval. This is due to the evidence (see previous discussion) that the value of
${\rm M_{core}^{He}}$ at the He-flash is fairly constant in the low-mass star range. However, ${\rm M_{core}^{He}}$ decreases for increasing metallicity, while the TRGB bolometric luminosity increases due to the increased efficiency of the H-shell, which compensates for the reduced core mass.

When considering the TRGB magnitude in a photometric system, the behaviour of the bolometric correction to the I-Cousin band (hereinafter, I-band) as a function of [Fe/H] and effective temperature compensates for the variation of the bolometric luminosity with metallicity at least for ${\rm [M/H]\le-0.9}$ (Lee et al.~1993). 
This occurrence stands from the evidence that  $\rm M_{bol}^{TRGB}$ is proportional to ${\rm \sim-0.18[M/H]}$, while $\rm BC_{I}$ is proportional to ${\rm \sim-0.14[M/H]}$. Therefore, the slope of the ${\rm BC_{I}-[M/H]}$ relationship is quite similar to the slope of  the $\rm M_{bol}^{TRGB}-[M/H]$ relationship, and being $\rm M_{I}^{TRGB}$=$\rm M_{bol}^{TRGB}- \rm BC_{I}$, it results that $\rm M_{I}^{TRGB}$ is almost independent of the metallicity: roughly speaking ${\rm M_I^{TRGB}}$ changes by less than 0.1 mag when the metallicity varies in the range ${\rm -2.2\le[M/H]\le-0.6}$, whereas  ${\rm M_V^{TRGB}}$ changes by $\sim2.4$~mag in the same metallicity range.

As far as it concerns the uncertainties affecting theoretical predictions about the luminosity of the TRGB , it is clear that, being the TRGB brightness fixed by the He core mass, any uncertainty affecting the predictions about the value of ${\rm M_{core}^{He}}$ immediately translates in an error on $\rm M_{bol}^{TRGB}$.

Theoretical analysis on this issue (Castellani \& Degl'Innocenti~1999, Salaris et al.~2002, Cassisi et al.~2007) have shown that the physical inputs which have the largest impact in the estimate of ${\rm M_{core}^{He}}$ are: atomic diffusion efficiency and conductive opacity evaluations.

{\it Atomic diffusion} -- to account for the occurrence of atomic diffusion causes an increase of  ${\rm M_{core}^{He}}$ as a consequence of the changes in the envelope opacity stratification: the  mass difference of the He core masses decreases as metallicity increases, from ${\rm \sim0.003M_\odot}$ at Z = 0.0001 to ${\rm 0.002M_{\odot}}$ at solar metallicity (Michaud et al.~2010). In the same time, a change by a factor of 2 in the efficiency of microscopic diffusion causes a change of about ${\rm (-0.002/+0.004)M_\odot}$ in the value of ${\rm M_{core}^{He}}$: the He core mass increasing when the atomic
diffusion efficiency is increased.

{\it Conductive opacities} -- since the conductive transport efficiency regulates the thermal state of the electron degenerate He core, a reliable estimate of the conductive opacities is fundamental for deriving the correct value of the He core mass at the He-flash. As rule of thumb, higher conductive opacities cause a less efficient cooling of the He-core and an earlier He-ignition, hence at a lower core mass.   
   
Until few years ago, only two choices for the conductive opacity were available, neither of which is totally satisfactory: the analytical relation provided by Itoh et al.~(1983, I83),  or the old Hubbard \& Lampe (1969, HL) tabulation.   
As pointed out by Catelan et al.~(1996),   the results by I83 were an improvement over the older HL ones, but their range of validity did not cover the He cores of RGB stars. When using the I83 conductive opacity, Castellani \& Degl'Innocenti~(1999) found an increase - with respect to the models based on the HL opacity - by 0.005${\rm M_{\odot}}$ of ${\rm M_{core}^{He}}$ core at the He-flash for a 0.8${\rm M_{\odot}}$ model with metallicity Z=0.0002, while in case of a 1.5${\rm M_{\odot}}$ star with solar chemical composition the increase amounts to 0.008${\rm M_{\odot}}$.
 
New conductive opacity estimates have been provided by Potekhin (1999). This set represents a significant improvement (both in the accuracy and in the range of validity) with respect to previous estimates: RGB stellar models based on these new conductive opacities provided He core masses at the He-ignition whose values are intermediate between those provided by the previous conductive opacity estimates (although more similar to the determinations based on the Itoh et al.~(1983) ones). However,  not even these conductive opacity are completely free of some relevant shortcomings. In this context, a significant improvement has been recently achieved with the updated evaluations provided by Cassisi et al.~(2007): this opacity set fully covers the thermal conditions characteristic of electron degenerate cores in low-mass, metal-poor stars, account for arbitrary chemical mixtures, and takes into account the important contribution from electron-ion scattering and electron-electron scattering in the regime of partial electron degeneracy (the most relevant case for the He cores of low-mass RGB stars).  
The new conductive opacities provide He-core masses at the He flash lower than in the case when the Potekhin (1999) calculations are used: the difference amounts to ${\rm 0.006M_\odot}$, regardless of the value of $Z$. The reason is that the new opacities are larger than the older ones, thus producing a different thermal stratification in the He core. The difference in He-core mass at the RGB tip causes a change of the RGB tip brightness: new models appear to be fainter by ${\rm \Delta{log(L/L_\odot)}\approx0.03}$.

Concerning the EOS, the analysis by Cassisi et al. (2003) shows that, when the adopted EOS accounts for all the various physical processes at work in the dense core of RGB stars, the residual uncertainty on the value of ${\rm M_{core}^{He}}$ can be small. 

The impact of current uncertainties on the relevant nuclear reaction rates as the one corresponding to the $3\alpha$ process has been recently investigated by Weiss et al.~(2005) with the result that current uncertainty has no significant influence on theoretical predictions about the TRGB.
However, there is an other nuclear reaction whose efficiency could affect the value of ${\rm M_{core}^{He}}$ core at the TRGB, i.e. the ${\rm ^{14}N(p,\gamma)^{15}O}$ nuclear reaction, via its impact on the shell H-burning efficiency. As already mentioned, the LUNA experiment has significantly improved the low-energy measurements of the reaction rate associated to this nuclear process. Weiss et al.~(2005) and Pietrinferni et al.~(2009) have shown that the LUNA reaction rate leads to an increase of ${\rm \sim0.002-0.003M_\odot}$ in the mass of the He core at the He-flash, with respect models based on the previous NACRE rate. 
Although the He-core mass is larger in the LUNA models, the surface luminosity at the TRGB is lower, the difference being ${\rm \Delta{log(L/L_\odot)}\approx0.02}$ dex. This is due to the lower efficiency of the CNO cycle in the H-burning shell, which compensates for the luminosity increase expected from the higher core mass.

\begin{figure}
\begin{centering}
\vskip -0.5cm
\includegraphics[width=7.5cm]{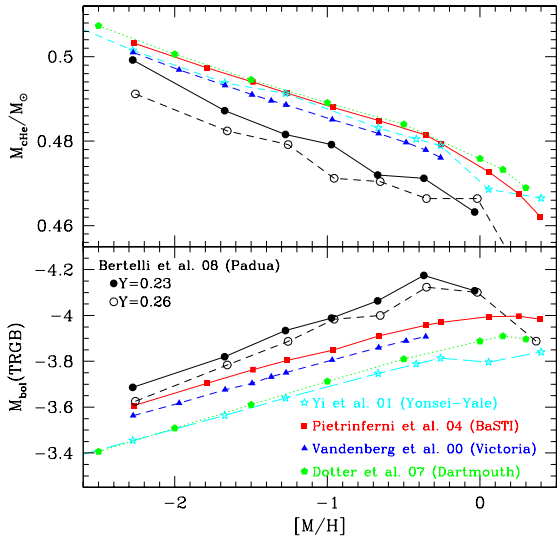}
\vskip -0.6cm
\caption{The trend of ${\rm M_{core}^{He}}$ and ${\rm M_{Bol}}$ at the RGB tip as a function of the metallicity from various stellar model libraries.
The displayed quantities refer to a 0.8${\rm M_{\odot}}$ model computed by adopting a scaled solar metal distribution.}
\end{centering}
\label{trgbmche}
\end{figure}

The comparison of recent results  (Bertelli et al. 2008 - Padua, Pietrinferni et al. 2004 - BaSTI, Vandenberg et al. 2000 - Victoria, Dotter et al. 2007 - Dartmouth, Yi et al. 2001 - Yonsei-Yale) concerning the TRGB bolometric magnitude and $\rm M_{core}^{He}$ at the He-flash is shown in fig.~16. When excluding the Padua models, there exists a fair agreement among the various predictions about  $\rm M_{core}^{He}$: at fixed metallicity the  spread among the various sets of models is at the level of $\rm 0.003M_\odot$. Concerning the trend of  $\rm M_{bol}^{\rm TRGB}$, all model predictions at a given metallicity are in agreement within $\sim0.15$ mag, with the exception of the Padua models that appear to be brighter, at odds with the fact that they predict the lowest $\rm M_{core}^{He}$  values. When neglecting the Padua and Yonsei-Yale models, the $\sim0.1$ mag   
spread among the different TRGB brightness estimates can be explained in terms of differences in the adopted physical inputs.  

\begin{figure}
\begin{centering}
\vskip -0.5cm
\includegraphics[width=7.5cm]{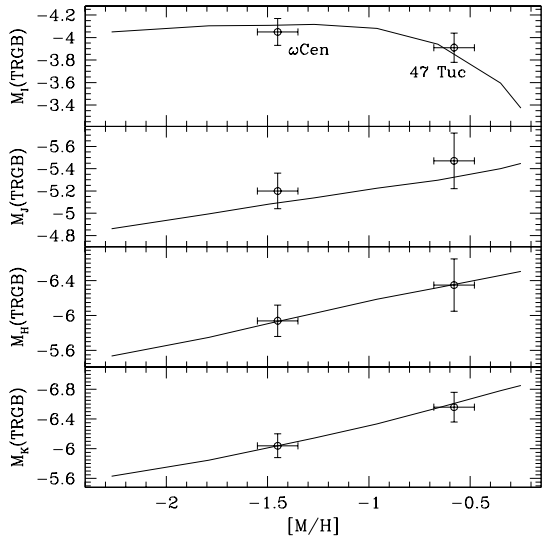}
\vskip -0.6cm
\caption{The magnitude of the TRGB in the I-band and in JHK near-infrared photometric filters as predicted by RGB models
based on the most updated physics (also concerning conductive opacity and nuclear cross sections; see text for more details). The empirical measurements of the absolute TRGB magnitude in $\omega$~Cen and 47~Tuc are also shown.}
\end{centering}
\label{trgbijhk}
\end{figure}

Due to the large relevance of the TRGB as standard candle, it is worthwhile showing a comparison between theoretical
predictions about the I-Cousin band magnitude of the TRGB and empirical estimates. This comparison is displayed in
the upper panel of fig.~17, where we show the measurements of the TRGB brightness for the two GGCs $\omega$~Cen
and 47~Tuc as provided by Bellazzini et al.~(2001, 2004). The same figure shows also the same kind of comparison but in the near-infrared JHK bands.
The model predictions shown in fig.~16 are based on the same physical framework adopted in the BaSTI stellar model library, but for the conductive opacity and the reaction rate for the ${\rm ^{14}N(p,\gamma)^{15}O}$ nuclear process for which the most updated evaluations provided by Cassisi et al.~(2007) and the LUNA experiment respectively, have been used. It is worth noting that the most updated TRGB models match very well empirical determination in all photometric filters, but in the J-band where the models appear to be slightly under luminous with respect the observations.

In this context, it can be useful to remember that, in order to derive a TRGB brightness calibration, a bolometric correction scale has to be used (Salaris \& Cassisi 1998) that, as it has been already mentioned, can be affected by large uncertainty. Therefore, it appears difficult at this time to disentangle the contribution to the global discrepancy between theoretical and empirical calibration, due to current uncertainty in the ${\rm BC_J}$ scale from that associated to present uncertainties in stellar RGB models. The evidence that the same theoretical calibration does not work perfectly for the J-band while providing a very good match in the I-Cousin band as well as in the other near-infrared bands shed light on the importance of an accurate and critical analysis on the uncertainties affecting the bolometric correction scales for the various photometric bands.

\section{The He-burning stage}
\label{hestage}

The Horizontal Branch (HB) is one of the most important evolutionary sequences in the CMD. There are many reasons for this relevance such as: 1) the brightness of the RR Lyrae stars and, more in general, the brightness of the HB, is the traditional distance indicator for metal-poor star clusters; 2) the number of stars observed along this branch enters in the R parameter definition (Buzzoni et al.~1983), the most important He indicator for old stellar systems; 3) the HB morphology is related to the decades-old issue of the \lq{second parameter}\rq\ in the GGC system.

From a theoretical point of view, although we know well for a long time the structural and evolutionary properties of HB stars,
we can not be yet fully confident in the theoretical predictions concerning this evolutionary phases, at least as far as it
concerns the luminosity and the evolutionary lifetime. This is simply due to the evidence that the evolutionary properties of HB
stars strongly depend on all the physical processes at work during the previous RGB phase. Therefore, the shortcomings affecting the physical scenario used for computing  H-burning structures appear, in some sense, amplified when considering HB stellar models which are, in addition, affected by other sources of uncertainty as the rates of the He-burning nuclear processes and the efficiency of mixing at the border of the convective core (Cassisi et al.~1998,~2003).

\begin{figure}
\begin{centering}
\vskip -1.5cm
\includegraphics[width=8cm]{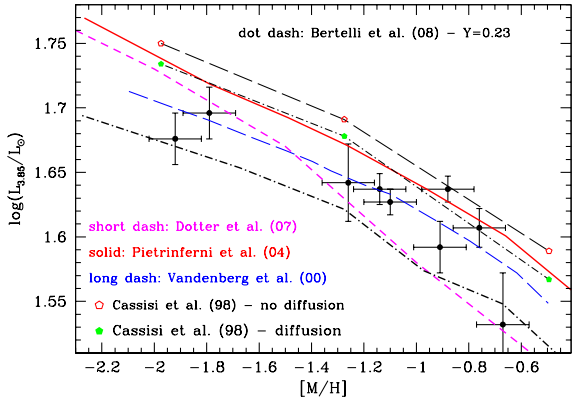}
\vskip -1.5cm
\caption{Comparison among various stellar model libraries concerning the ZAHB luminosity at ${\rm \log{T_{eff}}=3.85}$, takes as representative of the average effective temperature of the RR Lyrae instability strip. The points with error bars correspond to the semi-empirical estimates of the same quantity in selected GCCs (see De Santis \& Cassisi~1999).}
\end{centering}
\label{hbteo}
\end{figure}

Let us start discussing the level of reliability of theoretical predictions about the ZAHB brightness. 
The bolometric luminosity of a ZAHB structure is controlled by two parameters: mostly the He core mass and, to a minor extent, the chemical stratification of the envelope. On the basis of this consideration, one can easily realize that all the uncertainties affecting the value of ${\rm M_{core}^{He}}$ at the TRGB directly affect also the HB brightness. 
When considering only as sources of error, the atomic diffusion efficiency and the different choices about
the conductive opacity; one derives that the visual absolute magnitude of the HB is uncertain at the level of $\sim0.10$~mag. 
Figure~18 shows a comparison between some recent model predictions about the luminosity of ZAHB stellar structure at an effective temperature corresponding to the average ${\rm T_{eff}}$ of the RR Lyrae instability strip.

As for any evolutionary phase, the lifetime of the core He-burning phase - for a given total mass - does depend on the \lq{speed}\rq\ at which the nuclear processes occur, i.e., on the nuclear reaction rates, and on the amount of available fuel, i.e., in the case of burning occurring in a convective core, on the location of the outer convective boundary. Really, the uncertainty on the HB evolutionary lifetimes is dominated by the uncertainty on nuclear reaction rates and by the not-well known, efficiency of convective processes (see below).

Concerning the reaction rates,  it is worth emphasizing that the ${\rm {^{12}C}(\alpha,\gamma){^{16}O}}$ 
reaction is, together with the triple$-\alpha$ process, the most
important among those involved in the He-burning. This occurrence being due to the evidence that: i)
its nuclear cross-section strongly affects the C/O ratio in the core of carbon-oxygen white dwarfs and, 
in turn, their cooling times (Cassisi \& Salaris~2013); ii) when the abundance of He inside the convective core is significantly reduced, 
the ${\rm {^{12}C}(\alpha,\gamma){^{16}O}}$ reaction becomes strongly competitive with the $3\alpha$ reactions 
(which need three $\alpha$ particles) in supplying the nuclear energy budget. This means that the cross-section
of this nuclear process has a huge impact on the core He-burning lifetime as well as on the chemical stratification of the core at
the central He exhaustion.

Unfortunately, this reaction has a resonance and a very low cross-section at low energies, and so the
nuclear parameters are difficult to measure experimentally or to calculate by theoretical analysis. As a consequence,
its reaction rate (${\rm \sigma_{C+\alpha}}$) is still affected by  an uncertainty of a factor of 2 (Kunz et al~2002). As already mentioned the uncertainty on the ${\rm {^{12}C}(\alpha,\gamma){^{16}O}}$ reaction rate strongly affects the HB lifetime, more in detail, the HB lifetime change correlates with a variation of the rate as ${\rm \frac{\Delta{t_{HB}}}{t_{HB}}\sim0.10\frac{\Delta\sigma_{C+\alpha}}{\sigma_{C+\alpha}}}$. So an uncertainty of a factor of 2 on ${\rm \sigma_{C+\alpha}}$ implies that HB lifetimes could be affected by an indetermination at the level of $\sim20$\%.

The HB lifetime strongly depends also on the efficiency of convection-induced mixing at the boundary of the convective core: during the core He-burning stage, the treatment of mixing at the boundary of the convective core,  is really a relevant problem. In fact, as a consequence of the burning process, helium is transformed into carbon and oxygen whose associate opacity is larger with respect that of an He rich mixture. This change in the opacitive properties of the stellar matter in the core,
strongly modifies the behaviour of the radiative gradient, producing an increasing of the mass size of the convective core (see Salaris \& Cassisi~2005, and references therein).

Unfortunately, in spite of the many theoretical works published over the last three decades, the physics that 
determines the extent of this convective region is still poorly known. The theoretical calculations 
available so far leave various scenarios open. Stellar models  based on a bare Schwarzschild criterion are still computed, and widely used in many studies. However, models that include some algorithm  to handle the discontinuity of the opacity that forms at the external border of the convective core  as a consequence of the conversion of He into C and O should be considered as more reliable (Castellani et al. 1971a, 1971b; Demarque \& Mengel 1972; Sweigart \& Gross 1976; Dorman \& Rood 1993). According to this mixing scheme, the change in the opacitive properties of the core naturally leads to the growth of the convective core (the so-called induced overshoot) and to the formation of a semiconvective layer outside the fully convective region. In an alternative approach, it is assumed that a mechanical overshoot  takes place at the boundary of the convective region (Saslaw \& Schwarzschild~1965; Girardi et al.~2000).  Although, the real occurrence of this phenomenon is out of debate, a quantitative estimate of the overshoot efficiency is still an unsettled issue.

Near the end of the core He-burning phase, there is another process associated with mixing, that could potentially largely affect the evolutionary properties of the models: a sort of pulsating instability of convection, the so-called {\sl breathing pulse} (Castellani et al. 1985), can occur, driving fresh helium  into the core and so, affecting the core He-burning lifetime as well as the carbon/oxygen ratio at the center of the star. It is still under debate if this mixing instability occurs in real stars or, if it is a fictitious process occurring as a consequence of our poor treatment of convection, as for instance, of the commonly adopted assumption of instantaneous mixing.

The approach adopted for managing mixing processes in HB structures affects largely the evolutionary lifetimes
as a consequence of the change in the available amount of fuel, but in addition it largely affects also the C/O ratio in the CO core at the central He exhaustion. Therefore the effects are quite similar to different assumptions about the rate for the nuclear process ${\rm {^{12}C}(\alpha,\gamma){^{16}O}}$. Therefore, there exists a sort of degeneracy between this reaction rate and the efficiency of mixing during the core He-burning phase. 
However, the use of various, independent, empirical constraints, and their comparison with theoretical predictions can allow to disentangle the evolutionary and structural effects associated with nuclear reaction rates and mixing processes. In this context, it is very useful the analysis of star counts along both the HB and the following evolutionary stage, i.e. the Asymptotic Giant Branch (AGB), as well as the observational benchmarks provided by asteroseismological measurements.

During the last decade a large effort has been devoted to the calibration of the R parameter (Buzzoni et al. 1983) in order to estimate the primordial He abundance of the GGC system\footnote{The discovery of multiple populations in GGCs has revealed that these stellar systems can host stars characterised by distinct initial He abundances. Although as shown by Salaris et al. (2004), the average GC He abundance should not result very different with respect the primordial value, this observational evidence has significantly reduced the importance of the R parameter for estimating the primordial He abundance via the study of GGCs. This notwithstanding, this method is still very important for deriving some clue about the average He abundance in complex stellar system such as the Galactic bulge (Renzini 1994).}.
This parameter is defined as the number ratio between HB and RGB stars brighter than the HB. So its theoretical calibration is strongly affected by model predictions about the HB lifetime. The recent analysis performed by Cassisi et al.~(2003) and Salaris et al.~(2004) have shown that updated HB models based on the more recent evaluation of the ${\rm {^{12}C}(\alpha,\gamma){^{16}O}}$ rate and on the semiconvective mixing scheme\footnote{These models neglect also the occurrence of breathing pulses in the late phase of the core He-burning stage. For a detailed discussion about the reasons for which the
occurrence of this process in real stars is considered implausible we refer to Caputo et al.~(1989).} are able to provide an estimate of the initial He abundance in very good agreement with the measurements obtained through the analysis of the Cosmic Microwave Background anisotropies and primordial nuclesynthesis models. They also predict a value for the parameter $R_2$ (i.e. the ratio between the number of AGB stars and that of HB objects - Caputo et al.~1978)  in fair agreement with observations. In particular, one has to note that the $R_2$ parameter is strongly affected by the adopted mixing scheme: the larger the mixing during the core He-burning, the less the amount of He available for the subsequent AGB evolutionary phase (and hence the shorter the AGB lifetime).
Concerning asteroseismology, the analysis of non-radial pulsations in white dwarfs (e.g., Metcalfe et al.~2000,~2001) can provide important clues about the C/O ratio within the CO core as well as on the ratio between the CO core mass and the total WD mass. All these empirical constraints when analysed within a self-consistent, and updated evolutionary framework can be of extreme relevance in order to improve our knowledge on the physical processes at work in He-burning, low-mass stars.

\subsection{Non-canonical processes in HB stars}

\subsubsection{The observational scenario}

Although the general properties of core He-burning stars from both a structural and evolutionary point of view are well understood, we still lack of a sound interpretation of all the physical parameters definitively governing - together with the metallicity - the morphology of the HB in various stellar systems: the so-called \lq{second parameter problem}\rq. In addition to this evidence, in this last decade high-precision photometry and multi-objects, high-resolution spectroscopy of large sample of HB stars in GGCs revealed a number of anomalies that do not found a plain interpretation in the canonical evolutionary framework: the presence of narrow gaps, i.e. region along the HB showing a significant paucity of stars with respect both their blue and red boundaries; the fact that HB stars hotter than a critical ${\rm T_{eff}}$ appear brighter than expected on the basis of canonical models; the evidence that hot HB stars apparently show too low surface gravities; peculiar patterns in both the surface heavy element abundances and in the surface rotational velocities.

Nowadays, it appears clear that there exists a strong link between all these pieces of evidence, and that various physical mechanisms such as atomic diffusion, radiative levitation, rotation, mass loss occurring during both HB and RGB stages, and probably also non-canonical mixings (not accounted in standard stellar models) cooperate to produce the peculiarities observed in HB stars and their progeny. Even though significant improvements have been done in the stellar model computations, we still lack a sound, physically grounded, explanation of all the observational findings as a whole.
In the following we briefly review some of the most relevant observed anomalies and discuss the state-of-art in the modelling of the physical mechanisms responsible for these peculiarities (for an accurate review on both the observational and theoretical aspects of this issue we refer to Cassisi \& Salaris~2013).

Let us start with the so called \lq{Grundahl's jump}\rq: Str{\"o}mgren photometry of a large sample of GGCs (Grundahl et al.~1998,~1999) has shown the presence of a jump in the u-band photometry of HB stars. From a morphological point of view this jump is described as a systematic deviation in the u magnitude and/or the ${\rm (u-y)}$ color with respect to the canonical theoretical ZAHB. An interesting property of this phenomenon is that the jump is an ubiquitous feature observed in all GCs with an blue enough HB morphology. The ${\rm T_{eff}}$ value at which the u jump starts, is of the order of 11500~K, and it is independent on the GC metallicity. In addition, the size of the u jump is $\sim0.8$~mag in all GCs. One can also note that the observational data start approaching again the theoretical ZAHB for ${\rm T_{eff}\ge23000}$~K.

The discovery of the u jump is considered the most striking, indirect, evidence of the occurrence of diffusive processes in hot HB stars. More in detail, if an HB star is hot enough - this occurrence explains the existence of a limiting ${\rm T_{eff}}$ for the appearance of the  \lq{Grundahl's jump}\rq\ - radiative levitation changes the atmospheric abundances of metals; so causing an increase of the relative contribution of metals to the opacity with respect that provided by H. The change induced in the stellar spectrum would imply that a larger fraction of the flux would come out through the u filter: levitation of heavy elements decreases the far-UV flux and by back-warming increases the flux in the u passband.

The presence of chemical anomalies in the outer layers of hot HB stars was firstly predicted on theoretical ground by Michaud et al.~(1983) as a consequence of the possible occurrence of radiative levitation. 
However, only in recent times, this theoretical prediction was completely confirmed by direct empirical evidence of the existence of large abundance anomalies in HB stars in GGCs (Behr~2003, Pace et al.~2006, and references therein). The observational investigations have shown that, while in HB stars cooler than ${\rm T_{eff}\approx11000}$~K, the iron and other heavy element abundances are close to those corresponding to the GC initial chemical composition, hotter HB stars show remarkable enhancements of iron and other metal species. In particular, the majority of hot HB stars show iron abundances greater than the solar value. In the same time, while some chemical species such as Ti, Ca, P, Cr display large enhancements with respect the original values, there are other elements such as Mg and Si that, on the contrary, show very little, if any, enhancement.
Spectroscopic measurements of the surface He abundance in hot HB stars reveal also a significant under abundance of this element with respect the initial value: this occurrence is explained as the consequence of gravitational settling being at work in the envelope of these stars.
 
From an observational point of view, our knowledge of the rotational rate distribution of stars along the HB of GGCs has improved a lot  because in combination with the study of the chemical abundance anomalies in hot HB stars, accurate measurements of their project velocity ${v\sin{i}}$ have been also obtained (Behr~2003). 
The most significant result of these rotation measurements is that hot HB stars and HB objects cooler than ${\rm T_{eff}\approx11000}$~K have very different distributions of ${v\sin{i}}$ values, an occurrence that implies a large difference in the actual rotational rate. 

More in detail, it has been found that stars cooler than $\approx11000$~K can reach large values of $v\sin{i}$ as high as ${\rm (20-40)kms^{-1}}$ depending on the given GC; while hot HB stars show significantly lower rotational rates, limited to ${\rm \sim8~kms^{-1}}$.  As a consequence, it appears that hot and extremely hot HB stars are intrinsically slow rotators. It is worth mentioning that there are some outliers, i.e. hot HB stars with rotational rates larger than observed in other stars with similar ${\rm T_{eff}}$, and interestingly these objects do not show the metal abundance enhancements observed in HB stars with the same ${\rm T_{eff}}$. This occurrence suggests the existence of a link between rotational rate and efficiency of diffusive processes.

One can easily notice that there is a close similarity between the ${\rm T_{eff}}$ values in correspondence of which very different empirical findings appear such as the rotational rate dichotomy, appearance of the chemical abundance anomalies, and the u jump: an occurrence pointing out the existence of a deep link between the physical mechanisms causing these effects.

\subsubsection{The theoretical scenario}

The various observational findings, previously mentioned, point out that diffusive processes have to be at work efficiently in the atmospheres of hot HB stars. Indeed, it was early suggested by Greenstein~(1967) and Michaud et al. (1983) that the underabundance of helium in the atmosphere of field HB stars could be the consequence of He sinking under the influence of gravity from surface to the interior layers. However, in order this process to be efficient, the outer stellar layers have to be stable against convection because the presence of any mixing would completely erase the chemical gradient produced by atomic diffusion. In case the atmosphere is stable and the stellar surface is hot enough so that radiation pressure can efficiently transfer momentum to the chemical species which present sufficiently large cross-sections to the outgoing radiation field, radiative levitation can be also efficient, so producing photospheric enhancement of (some) heavy elements.

Therefore, a crucial property in defining the efficiency of diffusive processes is the mass of the outer convection zone, because it is thoroughly mixed and the chemical abundance gradients will appear below this mixed region. As shown in fig.19, HB structures cooler than ${\rm T_{eff}\approx6300}$~K have a very deep convective envelope, for ${\rm T_{eff}}$ larger than this limit but lower than $\sim11500$~K, the thin convective zones associated to H ionization and to the first ionization of He, lie at - or slightly below - the stellar surface. For ${\rm T_{eff}>11500}$~K, only the convective region due to He II is present and its location in mass moves outward with increasing ${\rm T_{eff}}$, until it approaches the stellar surface at ${\rm T_{eff}\sim23000}$~K.

It is the disappearance of convection at ${\rm T_{eff}\approx11500}$~K that controls the appearance of the effects of diffusive processes. In this context, the model predictions about the dependence of outer convective zones on the effective temperature provide a self-consistent interpretation of the reason for which the observed chemical abundance anomalies, and the u jump appear at - almost - the same ${\rm T_{eff}}$ value.

\begin{figure}
\begin{centering}
\vskip -.5cm
\includegraphics[width=9cm]{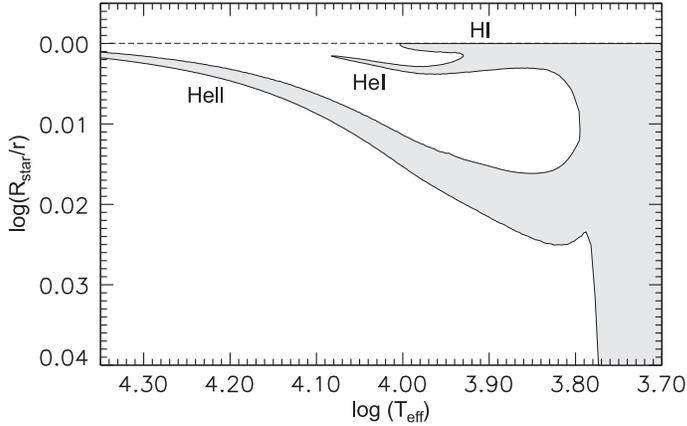}
\vskip -.2cm
\caption{The location with respect the stellar surface of the convective zones (shaded areas) in canonical ZAHB stars as a function of the their ${\rm T_{eff}}$. The convective zones associated with H~I, He~I and He~II  ionization zones - when separated -are clearly marked.}
\end{centering}
\label{conv}
\end{figure}

In this last decade, a huge theoretical effort (Michaud et al.~2007,~2008,~2010) has been devoted to account for diffusive processes (both atomic diffusion and radiative levitation) in stellar model computations. These models have been very successful in explaining the physical processes occurring in the atmosphere of hot HB stars as well as in interpreting the selective heavy element enhancements induced by radiative levitation (some chemical elements would be more affected by radiative forces than other ones). 

However, these evolutionary models show that the predicted heavy elements enhancements are too large with respect spectroscopical measurements, and metal enhancements are also expected for stars cooler than 11500~K, in stark contrast with spectroscopical measurements. This means that the convective zones eventually present in canonical models are not able to properly limit the diffusive process efficiency. 

In present generation of diffusive HB models this problem is overcame by accounting for some {\sl ad hoc} (additional) mixing process (some kind of turbulence) operating in the outermost layers. So far we have no clue about the physical meaning of the mechanism(s) operating in real stars which partially inhibit the occurrence of diffusive processes. 

Actually, various hydrodynamic processes may compete with diffusive processes and reduce their impact on surface chemical abundances. Among the various processes a pivotal role could be played by: mass loss, rotationally-induced meridional circulation, turbulent diffusion. The discussed observational evidence about the rotational properties and chemical anomalies observed in HB stars seems to suggest that rotation and the correlated mixings could be the process competing with diffusion

A link between rotational rate distribution and chemical anomalies has been recently suggested by Quievy et al.~(2009). In canonical modes, due to gravitational settling, the envelope He abundance becomes so low that the convection zone associated with He ionization disappears. When this occurs, diffusive processes can hugely modify the surface chemical stratification being their effects not halted by convection. However, if the star rotates with a rotational rate larger than some critical (still unknown) value, the rotationally-induced meridional circulation below the outer convective zone could prevent atomic diffusion to reduce the envelope He content below the critical abundance for which the He convection zone disappears.

Preliminary analysis have shown that the observed rotational velocities of HB stars hotter than $\sim11500$~K are always below the critical velocity so rotationally-induced mixing cannot compete with He settling, an occurrence that produces the disappearance of the He convective zone, and then the concomitant development of the observed chemical anomalies. On the other hand, stars cooler than ${\rm T_{eff}\approx10500}$~K show a rotational velocity larger than the critical velocity estimate and then in such structures meridional circulation would prevent the diffusive processes to significantly alter the chemical stratification in the outer layers as indeed observed. This link between rotation and diffusive processes would provide an explanation of the reason for which diffusive HB models - that at this stage do not yet account for rotationally-induced mixing -  predict large metal enhancements also in cool stars. A more accurate evaluation of the critical velocity for efficient rotationally-induced mixing, as well as of the occurrence of this non canonical mixing are mandatory before fully settling the evolutionary scenario.

The empirical evidence concerning the rotational rate distribution of HB stars in GCs is somehow more difficult to be explained than the observed chemical anomalies, and quite more challenging for the theoretical evolutionary framework. 
The main problems in interpreting the observational findings are related to the existence of a dichotomy in the rotational rate distribution with some objects showing high rotational rates, and to the fact the bluest stars are the more slow-rotators among HB stars.

 In fact, following early suggestions (Mengel \& Gross~1976) it has been always considered that rotating stars would delay the ignition of He at the RGB tip so attaining a larger luminosity and a lower ${\rm T_{eff}}$ with respect non-rotating objects. All these effects would contribute to enhance the mass loss during the RGB stage, from this behaviour one could expect that the larger the rotational rate in the RGB stage, the lower the envelope mass of the corresponding HB stars and, hence, the hotter is the ZAHB location.
Such \lq{naive}\rq\ expectation is in apparent contrast with observations showing that the more hot HB stars are indeed the slowest rotators along the HB. The situation is not very clear because mass loss actually removes also angular momentum, so bluer HB stars are expected to have a lower angular momentum  than redder HB stars, but having quite smaller radii they have also a quite smaller moment of inertia. Therefore, it is not possible to predict a priori if hot HB stars should show lower rotational rate. 
Another embarrassment arises also from the mere existence of fast HB rotators: due to the large mass loss during the RGB stage, huge amount of angular momentum should be lost during this evolutionary stage, so it is not clear the source of the angular momentum forcing the envelope of some HB stars cooler of $\sim11500$~K to rotate as fast as observed.

So far there are no many theoretical investigations devoted to study the evolution of the rotational rate when a star moves from the MS up to the RGB tip and then settles on the ZAHB. The results provided by Sills \& Pinsonnealt (2000) show that, under specific - but still
acceptable - assumptions about the initial angular momentum distribution, and allowing a redistribution of angular momentum from the - faster -  rotating core to the outer envelope, then fast rotating HB stars with rotational velocities comparable to the faster rotating HB stars in GCs, can be predicted. However, these models are not able to predict at all the extremely slow rotation rates observed in hot HB stars. 

Several different solutions have been proposed in order to make inefficient, in hot HB stars, the transport of angular momentum from the stellar core to the envelope, such as: high mass loss efficiency during the core He-burning stage (Vink \& Cassisi~2002), and the presence of a huge mean molecular weight gradient originated by the gravitational settling of He (as indeed occurs in HB stars hotter than the Grundahl's jump).

\section{Final remarks}

The reliability and accuracy of the theoretical evolutionary framework improved significantly during last decade. A proof of this is represented by the capability of last generation of stellar models to reproduce fairly well many observational constrains such as CMD of resolved stellar populations and empirical data for
well-studied binary systems. This notwithstanding, also the most updated stellar models are not completely free from uncertainties and shortcomings,
a clear proof being the occurrence of - sometimes not-negligible - differences between results provided by independent research groups. 

From the point of view of stellar models users, the best approach to be used for properly accounting for these uncertainties, is to not use evolutionary results with an uncritical approach, and also to adopt as many as possible independent theoretical predictions in order to have an idea of the uncertainty existing in the match between theory and observations. It would be also worthwhile to devote some attention to understand the improvements adopted both in the physical inputs as well as in the physical assumptions, by people computing the evolutionary models.

On the other hand, stellar model makers should continue their effort of continuously updating their models in order to account for the \lq{best}\rq\ physics available at any time, and consider the various empirical constraints as a benchmark of their stellar models. 
The painstaking effort to compare the evolutionary framework with observations is mandatory in order to have a feedback on the accuracy of the adopted physical framework, and represents a crucial step for obtaining as much as possible accurate and reliable stellar models.

In the previous sections, we have mentioned that a quite important source of uncertainty in the comparison between theory and observations arises from the errors still affecting both theoretical and empirical color - effective temperature relations and the bolometric correction scales for the various photometric bands. It is evident that these uncertainties strongly hamper the possibility of a sound comparison between models and empirical evidence and, of course, make extremely problematic to assess the level of accuracy of present evolutionary scenario. In our belief, a big effort should be devoted in the near future to improve the reliability of the transformations adopted for transferring stellar models from the H-R diagram to the various observational planes.

It has also to be emphasised that the uncertainties on both GGCs metallicity and distance scale strongly hamper the possibility to realise a meaningful
comparison between theory and observations. Although, large improvements have been achieved in these fields, current errors are still too large for
offering the opportunity of a plain assessment of residual uncertainties in the evolutionary scenario

On the basis of these considerations, it appear clear that large improvement in the stellar evolution framework could be achieved in the near future only if scientists working in various fields of research, from fundamental physics to observational astrophysics, will provide their own contribution to improve our description of the \lq{physics}\rq\ at work in real stars.

\acknowledgements           
I wish to warmly acknowledge M. Salaris and A. Pietrinferni for the very pleasant and stimulating collaboration all along many years.
It is also a real pleasure to thank C. Charbonnel, Y. Lebreton and D. Valls-Gabaud for organising this interesting school, for all the help provided, and very pleasant discussions. This research has made use of NASA's Astrophysics Data System Bibliographic Services.


\end{document}